\documentclass{emulateapj}

\newcommand{\snuB}{$0.35$}

\newcommand{\snuBstat}{$^{+0.17}_{-0.12}$}
\newcommand{\snuBsys}{${\pm0.01}$}
\newcommand{\snuBclass}{${\pm0.13}$}
\newcommand{\snuM}{$0.112$}

\newcommand{\snuMstat}{$^{+0.055}_{-0.039}$}
\newcommand{\snuMsys}{${\pm0.005}$}
\newcommand{\snuMclass}{${\pm0.042}$}
\newcommand{\snuBO}{$0.68$}

\newcommand{\snuBOstat}{$^{+0.34}_{-0.24}$}
\newcommand{\snuBOsys}{${\pm0.03}$}
\newcommand{\snuBOclass}{${\pm0.26}$}

\begin{document}

\shorttitle{The SN Ia Rate in $0.5<z<0.9$ Galaxy Clusters}
\shortauthors{Sharon et al.}

\title{The Type Ia Supernova Rate in Redshift 0.5--0.9 Galaxy Clusters\altaffilmark{*}}
\altaffiltext{*}{Based on observations made with the NASA/ESA 
  {\it Hubble Space Telescope}, obtained at the Space Telescope Science
  Institute, which is operated by the Association of Universities for
  Research in Astronomy, Inc., under NASA contract NAS 5-26555. These
  observations are associated with programs GO-10493 and GO-10793, and with 
archival programs GO-9033, GO-9090, GO-9290, GO-9292, GO-9722, GO-9744, GO-9836,  and
GO-10509.
  Based in part on data collected at the Subaru telescope, which is operated 
  by the National Astronomical Observatory of Japan. 
  Some of the data
  presented herein were obtained at the W. M. Keck Observatory, which
  is operated as a scientific partnership among the California
  Institute of Technology, the University of California, and NASA; it was
  made possible by the generous financial support of the W. M. Keck
  Foundation.}
\author{Keren Sharon\altaffilmark{1,2}, 
{Avishay Gal-Yam\altaffilmark{3}}, 
{Dan Maoz\altaffilmark{2}},
{Alexei V. Filippenko\altaffilmark{4}, Ryan J. Foley\altaffilmark{4,10,12}, 
Jeffrey M. Silverman\altaffilmark{4}},
{Harald Ebeling\altaffilmark{5}, Cheng-Jiun Ma\altaffilmark{5}},
{Eran O. Ofek\altaffilmark{6}},
{Jean-Paul Kneib\altaffilmark{7}},
{Megan Donahue\altaffilmark{8}},
{Richard S. Ellis\altaffilmark{6}},
{Wendy L. Freedman\altaffilmark{9}},
{Robert P. Kirshner\altaffilmark{10}},
{John S. Mulchaey\altaffilmark{8}},
{Vicki L. Sarajedini\altaffilmark{11}}, and
{G. Mark Voit\altaffilmark{8}}}

\altaffiltext{1}{Kavli Institute for Cosmological Physics, The University of Chicago, Chicago, IL 60637.}
\altaffiltext{2}{School of Physics and Astronomy, Tel Aviv University, Tel Aviv 69978, Israel.}
\altaffiltext{3}{Benoziyo Center for Astrophysics, Faculty of Physics, Weizmann Institute of Science, Rehovot 76100, Israel.}
\altaffiltext{4}{Department of Astronomy, University of California, Berkeley, CA 94720-3411.} 
\altaffiltext{5}{Institute for Astronomy, University of Hawaii, 2680 Woodlawn Drive, Honolulu, HI 96822.}
\altaffiltext{6}{Division of Physics, Mathematics, and Astronomy, California Institute of Technology, Pasadena, CA 91125.}
\altaffiltext{7}{Laboratoire d'Astrophysique de Marseille, CNRS-Université Aix-Marseille, 38 rue F. Joliot-Curie, 13388 Marseille Cedex 13, France.}
\altaffiltext{8}{Department of Physics and Astronomy, BPS Building, Michigan State University, East Lansing, MI 48824.}
\altaffiltext{9}{Carnegie Observatories, 813 Santa Barbara Street, Pasadena, CA 91101.}
\altaffiltext{10}{Harvard-Smithsonian Center for Astrophysics, 60 Garden Street, Cambridge, MA 01238.}
\altaffiltext{11}{Department of Astronomy, University of Florida, Gainesville, FL 32611.}
\altaffiltext{12}{Clay Fellow.}

\begin{abstract}\label{s.abstract}

Supernova (SN) rates are potentially powerful diagnostics of metal
enrichment and SN physics, particularly in galaxy clusters with their deep, 
metal-retaining potentials and relatively simple star-formation histories.
We have carried out a survey for supernovae (SNe) in galaxy clusters,
at a redshift range $0.5<z<0.9$, using the Advanced Camera for Surveys
(ACS) on the {\it Hubble Space Telescope}.  We reimaged a sample of
15 clusters that were previously imaged by ACS, thus obtaining two to
three epochs per cluster, in which we discovered five likely cluster
SNe, six possible cluster SNe~Ia, two hostless SN candidates, and
several background and foreground events. Keck spectra of the host
galaxies were obtained to establish cluster membership.  We conducted
detailed efficiency simulations, and measured the stellar luminosities
of the clusters using Subaru images. We derive a cluster SN rate of
\snuB~SNu$_B$\, \snuBstat ~(statistical)\, \snuBclass ~(classification)\,
\snuBsys ~(systematic) [where SNu$_B$ = SNe (100\, yr $10^{10}\,
  {\rm L}_{B,\sun}$)$^{-1}$] and \snuM~SNu$_M$\, \snuMstat
~(statistical)\, \snuMclass ~(classification)\, \snuMsys ~(systematic)
[where SNu$_M$ = SNe (100\,yr\, $10^{10}\, {\rm M}_{\sun}$)$^{-1}$]. 
As in previous measurements of cluster SN rates, the uncertainties 
are dominated by small-number statistics.
The SN rate in this redshift bin is consistent with the SN rate in
clusters at lower redshifts (to within the uncertainties), and shows
that there is, at most, only a slight increase of cluster SN rate with
increasing redshift.  The low and fairly constant SN~Ia rate out to $z 
\approx 1$ implies that the bulk of the iron mass in clusters was already 
in place by $z\approx 1$. The recently observed doubling of iron abundances 
in the intracluster medium between $z=1$ and $0$, if real, is likely the
result of redistribution of existing iron, rather than new production of
iron.

\end{abstract}


\section{Introduction}\label{s.introduction}

Quantifying the rates and properties of supernovae (SNe) in
high-redshift galaxy clusters is important for several applications.
In structure-formation studies, SNe play a crucial role in baryonic
physics.  Their energy deposition into the environment is relevant to
both galaxy formation and star formation. Numerical simulations of
galaxy formation now include feedback from SN explosions (e.g.,
Borgani et al. 2004; Kay et al. 2007; Nagai et al. 2007; Scannapieco
et al. 2008; see Borgani et al 2008a,b for reviews), but the
efficiency of this feedback is unknown.

In terms of cosmic metal-enrichment history, SNe are the sources of
iron and other heavy elements that can be observed in the intracluster
medium (ICM) and are detectable through X-ray observations (e.g.,
Balestra et al. 2007; Maughan et al. 2008; de Plaa et al. 2007). The
abundances of these elements in the ICM depend on the integrated
history of SN explosions (e.g., Maoz \& Gal-Yam 2004), as all of the
elements produced during all stages of cluster formation and evolution
must remain in the cluster due to its deep potential well. The
abundances also depend on the efficiency with which matter is ejected
from galaxies into the ICM, whether by SN-driven galactic winds (De
Young 1978; White 1991; Renzini 1997; Borgani et al 2008b; Sivanandam
et al. 2009), by gas stripping due to ram pressure (Gunn \& Gott 1972;
Mori \& Burkert 2000), or by galaxy-galaxy interactions (e.g., Clemens
et al. 2000).  SNe from a diffuse intergalactic stellar population may
also be non-negligible contributors to the ICM enrichment (Gal-Yam \&
Maoz 2000a, 2000b; Gal-Yam et al. 2003; Lin \& Mohr 2004; Tornatore et
al. 2007).  Measuring the properties and rates of SNe of all types in
clusters as a function of redshift can thus shed light on galaxy and
cluster formation.

Finally, cluster SN rates can provide clues for our understanding of
Type Ia supernova (SN~Ia) physics.  It is widely agreed that SNe~Ia
are the thermonuclear explosions of near-Chandrasekhar-mass
carbon-oxygen white dwarfs in binary systems. However, the
nature of the progenitor systems is still not known, and several
different channels have been proposed (see, e.g., Mannucci et al. 2008
for a recent overview).  One prediction of a progenitor scenario that
can be tested by observations is the delay-time distribution (DTD)
between formation of a stellar population and the SN~Ia explosion of
some of its members. In recent years, constraining the DTD has been
attempted by comparing cosmic star-formation history (SFH) to
redshift-dependent rates of SNe~Ia in the field (e.g., 
Gal-Yam \& Maoz 2004; Dahl{\'e}n et
al. 2004, 2008; Cappellaro et al. 2005; Neill et al. 2006; Botticella
et al. 2008; Poznanski et al. 2007; Kuznetsova et al. 2008).
A major complication in such measurements is the
observational uncertainty in the SFH (e.g., F{\"o}rster et
al. 2006). In a further recent development, several studies have found
evidence for the coexistence of two SN~Ia explosion channels, a
``prompt'' channel that leads to an explosion within $\sim10^8$ yr of
the formation of a progenitor binary system, and a ``delayed'' one that
occurs at least several Gyr after star formation and dominates in old stellar
environments (Mannucci et al. 2005, 2006; Scannapieco \& Bildsten
2005; Sullivan et al. 2006b; Totani et al. 2008; Pritchet et al. 2008;
Aubourg et al. 2008; Raskin et al. 2009; Maoz et al. 2010; Brandt et al.
2010; Maoz \& Badenes 2010). 

Galaxy clusters form unique environments for DTD studies.
The  stellar population in galaxy clusters is dominated by old stars
in early-type galaxies, particularly in the core of the cluster (e.g.,
Visvanathan \& Sandage 1977; Renzini 2006) with a
very small amount of star formation taking place, mainly in
star-forming galaxies at the outskirts of the clusters
(e.g., Hansen et al 2009; Porter et al. 2008; Bai et al. 2007;
Saintonge et al. 2008; Loh et al. 2008). The fraction of star-forming galaxies increases with redshift
(Butcher \& Oemler 1978, 1984), an effect that is independent of cluster richness (Hansen et al 2009).
Since clusters have little
ongoing star formation, measuring the redshift-dependent SN~Ia rate in
clusters can isolate the delayed channel.  Moreover, since the SFH in
clusters is relatively simple compared to field galaxies, the DTD that can be deduced from the SN
rate depends less strongly on the details of the assumed SFH (Maoz \&
Gal-Yam 2004).

To date, the SN~Ia rate as a function of redshift in clusters has not been well
measured.  Until recently, the few existing published rates relied on
small numbers of detected SNe, and the large uncertainties were
dominated by small-number statistics.  An intermediate/high-$z$ SN~Ia
rate was derived by Gal-Yam et al. (2002) using archival 
{\it Hubble Space Telescope (HST)} imaging of 9 clusters, in which they
discovered two or three likely cluster SNe. Their measured rates were
$0.39^{+0.59}_{-0.25}$ and $0.80^{+0.92}_{-0.40}$ SNu$_B$ at $z=0.25$
and $z=0.9$, respectively, where SNu$_B$ denotes SNe (100\, yr\,
$10^{10}{\rm L}_{B,\sun})^{-1}$. 
These rates correspond to roughly $0.11^{+0.16}_{-0.07}$ 
and $0.22^{+0.25}_{-0.11}$  SNu$_M$, respectively, where SNu$_M$ 
denotes SNe (100\, yr\, $10^{10}\, {\rm M}_{\sun}$)$^{-1}$ 
(see \S~\ref{s.discussion}). Based on three cluster SNe and three 
possible cluster SNe, Graham et al. (2008) derived a SN~Ia
cluster rate at $0.2<z<1.0$ from the CFHT Supernova Legacy Survey 
(SNLS) of $0.1^{+0.09}_{-0.04}$ SNu$_M$. Mannucci et al. (2008) 
have reanalyzed the Cappellaro et al. (1999) nearby SN sample 
to derive a local ($z<0.04$) SN~Ia cluster rate, based on 11 SNe, of
$0.066^{+0.027}_{-0.020}$ SNu$_M$, which they found to be
significantly higher than the corresponding rate in field elliptical
galaxies,  $0.019^{+0.013}_{-0.008}$ SNu$_M$. Sharon et al. (2007) reported a rate of
$0.098^{+0.059}_{-0.039}\pm{0.009}$ SNu$_M$ at a slightly higher
redshift, $0.06<z<0.19$, based on the Wise Observatory Optical
Transient Survey (WOOTS) detection of six cluster SNe~Ia (Gal-Yam et
al. 2008). 
The Sloan Digital Sky Survey-II (SDSS-II)
Supernova Survey has discovered thousands of SN candidates, and by cross
correlation of the confirmed SNe with SDSS cluster catalogs (Koester et al. 2007, Miller et al. 2005)
measured cluster rates of 
$0.060 ^{+0.027 +0.002}_{-0.020 -0.001}$ SNu$_M$ at  $z=0.084$ and 
$0.088 ^{+0.022 +0.003}_{-0.018 -0.002}$ SNu$_M$  at  $z=0.225$  (Dilday et al. 2010). 

\begin{figure}[]
\rotatebox{0}{\scalebox{1}{\plotone {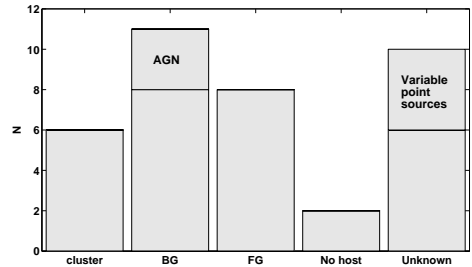}}}
\caption{Candidate classification. ``BG'' and ``FG'' denote 
background and foreground events, respectively.}
\label{fig.bars}
\end{figure}

Measurements of SN~Ia cluster rates from larger surveys are
ongoing. To name a few,
the Palomar Transient Factory (Law et al. 2009; Rau et al. 2009)
is expected to discover thousands of SNe in a footprint of $> 8000$ deg$^2$,
which will also be used to measure cluster SN rates at low redshift.
 At very high redshifts, the
Supernova Cosmology Project (SCP, PI Perlmutter) has targeted 25
clusters with a 219-orbit {\it HST} multi-epoch program (GO-10496)
in which about eight cluster SNe have been discovered (Dawson et al. 2009;
Melbourne et al. 2007), permitting derivation of the cluster
SN rate at $z \gtrsim 1$. 

Here we present results of an {\it HST}-based SN survey in galaxy
clusters at $0.5<z<0.9$.  Throughout the paper we assume a flat
cosmology, with parameters $\Omega_{\Lambda} = 0.7$, $\Omega_{m} =
0.3$, and $H_0 = 70$ km s$^{-1}$ Mpc$^{-1}$.  Magnitudes are reported
in the Vega-based system unless stated otherwise.


\section{{\it HST} Observations and Reductions}\label{s.survey}

We were allocated 30 {\it HST} orbits to reimage 15 high-redshift
galaxy clusters, during two observation cycles (Programs GO-10493 in
Cycle 14 and GO-10793 in Cycle 15, PI A. Gal-Yam). The clusters were
selected to be X-ray bright, in the redshift range $0.5<z<0.9$, and to
have been imaged with the Advanced Camera for Surveys (ACS) in the
past. We also required that archival data for each target were
non-proprietary at the time of our imaging, to ensure prompt detection
of SN candidates.  To enable comparison to archival data, the new
images were obtained using the same filters as the archival ones,
either the F814W filter ($\sim I$ band) or the F775W filter ($\sim i$ 
band). New observations were obtained at the same position angles as the 
archival ones, or rotated by $90^\circ$, $180^\circ$, or $270^\circ$, to 
allow maximal overlap between the images.  
The archival data consist of observations
from several {\it HST} programs; see Table~\ref{table.clusters2},
which also lists the other observations that were used in this survey,
and the area of overlap between epochs. Exposure times were typically
one orbit ($\sim2000$~s) per observing epoch in programs GO-10493,
GO-10793, GO-9292, GO-9744, and GO-9836, and at least two orbits
($\sim4500$~s) in programs GO-9033, GO-9090, GO-9290, GO-9722, and
GO-10509.  The $5\sigma$ detection limiting magnitudes for a point
source are typically $\sim26.4$ mag in the $I$ band and $\sim26.9$
mag in the $i$ band in our single-orbit observations, and $\sim27.5$
mag in the two-orbit archival $V$-band (F555W) images that exist for
some of the fields. 
The precise detection probabilities as a
function of magnitude for variable point sources are measured through
simulations, as detailed below in \S~\ref{S.efficiency2}.

We were able to obtain high-quality data for all but one cluster in
the original list during Cycle 14. The missing cluster,
MACSJ0025-1222, was withdrawn from our survey since it had not yet
been successfully observed (GO-9722) at the time of our survey. It was
replaced in our Cycle 15 target list with another cluster,
SDSSJ1004+4112, that had already been observed twice by {\it HST}
(Cycle 12, GO-9744; Cycle 14, GO-10509) and fits our selection
criteria. During Cycle 15, the halt in operations of ACS caused the early
termination of our program; six of the targets were not
imaged. We therefore have three epochs for nine clusters, and only two
epochs for six clusters.


Each new epoch was split into four dithered subexposures.  The
subexposures of each epoch were reduced using the standard
{\it HST}/ACS pipeline, and combined using the Multidrizzle routine
(Koekemoer et al. 2002) to remove cosmic-ray hits, dead or hot pixels,
and other artifacts such as trails from satellites that crossed the
field of view, with a square kernel, \texttt{pixfrac$=$1.0} and \texttt{scale$=$INDEF}.
Images from consecutive epochs of the same field were
then aligned and subtracted from each other, to form a difference
image. Specifically, transient candidates were searched
for in Epoch I compared to Epoch II, in Epoch II compared to Epoch I, and
in Epoch III compared to Epoch II. 
We note that point-spread function (PSF) matching was not
required (see \S~\ref{S.efficiency2} for more details).

Each difference image was searched by eye promptly after the
observation, and all transient or variable candidates were noted.
Obvious variable stars and known active galactic nuclei (AGNs) were
removed from the follow-up list at this stage.  Table
\ref{table.cands} lists the remaining candidates. 
Assuming that SN production follows light (e.g., Forster \& Schawinski 2008),
a galaxy was
considered to be a candidate's host if the candidate was within its
$2\sigma$ isophotal contour, defined as the contour along which the
galaxy flux per pixel is $2\sigma$ above the background fluctuations,
which in practice means the candidate is seen clearly embedded in the
galaxy light.  From the areas enclosed by these isophotal contours,
the probability for a chance association in a given image is $<2\%$.
Three of the 37 candidates do not satisfy this criterion, yet they
were also chosen as likely candidate-host associations due to their
small projected distances to their putative hosts, less than twice the
radius of the $2\sigma$ isophotal contour. In these cases, the
probability for a chance association is $6\%$.  When more than one
galaxy could be a likely host, follow-up spectroscopy was scheduled
for the additional galaxies as well.

As detailed below and shown in Figure~\ref{fig.bars}, among the 37 candidates, 
at least 6 are likely cluster events based on their host-galaxy redshifts
(Figure~\ref{fig.cluster_tiles}), 5 of which are likely SNe~Ia,  and 
one is likely a core-collapse (CC) SN. Two candidates have ambiguous
hosts and are possible cluster SNe~Ia.  Among the other candidates
(Figure~\ref{fig.field_tiles1}), 8 are background transients (BG), 8 are
foreground events (FG), and 3 proved to be AGNs. Two candidates have no
apparent host, and are possible cluster SNe~Ia.   The host galaxies of
the remaining 8 candidates are not confirmed by spectroscopy, and their
classification is also discussed below. Of these 8 unconfirmed
candidates, 4 are possible cluster SNe~Ia.  The remaining 4 were
variable (rather than transient) point sources that were detected at
all epochs and are not clearly associated with any host galaxy, and
are therefore probably quasars or Galactic variable stars.

\begin{deluxetable*}{p{2.25cm}p{0.5cm}p{1.45cm}p{1.5cm}p{0.8cm}p{0.1cm}p{0.9cm}p{0.1cm}p{0.2cm}p{0.3cm}p{0.3cm}p{0.8cm}p{0.3cm}p{0.3cm}p{0.3cm}p{0.3cm}p{0.3cm}}
\tablewidth{0pt} 
\tablecaption{Cluster Fields \label{table.clusters2}}
\tabletypesize{\scriptsize}
\tablecolumns{17} 
\tablehead{ 
           \colhead{Cluster }                                    &
           \colhead{ $z^b$}                             &
           \multicolumn{2}{c}{Coordinates (J2000)}     &
	  \multicolumn{2}{c}{Epoch I$^c$}& 
	  \multicolumn{5}{c}{Epoch II (GO-10493)} & 
	  \multicolumn{5}{c}{Epoch III (GO-10793)}&
	   \colhead{Filter}                   \\
	   \colhead{}                                       &
           \colhead{}                                       &
           \colhead{RA}                                       &
           \colhead{Dec}                                       &
           \colhead{date}                                       &
           \colhead{GO}                                       &
           \colhead{date}                                       &
           \colhead{area$^e$}                               &
           \colhead{$\Delta t^d$}                        &
          \colhead{L$_B^g$}&
          \colhead{M$^h$}&
           \colhead{date}                                       &
           \colhead{area$^e$}                                       &
          \colhead{$\Delta t^f$}                     &        
          \colhead{L$_B^g$}&
          \colhead{M$^h$}&
           \colhead{}                                       }           
\startdata
MACSJ2214$-$1359     	&0.503 	&22 14 57.34 	&$-$14 00 12.2 &2003/10 &9722 &2005/08        & 8.34  &100.5&4.4 &14.2 &\nodata & 0        &0       &0	&0         &F814W \\
MACSJ0911$+$1746     	&0.505 	&09 11 11.18 	&$+$17 46 34.8 &2004/03 &9722 &2005/10        & 8.44 &93.4  &2.2 &6.9   &2006/12 & 10.3  &101.2 &2.5    &7.6     &F814W\\
MACSJ0257$-$2325	&0.505 	&02 57 08.83 	&$-$23 26 03.3 &2004/01 &9722 &2005/08        & 8.05  &75.4  &2.1 &6.4   &2006/08 & 10.5  &99.0  &2.4    &7.5      &F814W \\
MS0451.6$-$0305$^a$  &0.538 	&04 54 10.48 	&$-$03 01 38.5 &2004/01 &9836 &2005/07        & 7.96  &83.3  & 3.0 &9.4  &\nodata & 0         &0      &0       &0	     &F814W \\
MACSJ1423$+$2404 	&0.543 	&14 23 48.60 	&$+$24 04 49.1 &2004/01 &9722 &2006/03        & 9.13  &94.4  & 3.1 &9.6  &\nodata & 0        &0       &0       &0	     &F814W \\
MACSJ1149$+$2223 	&0.544 	&11 49 35.51 	&$+$22 24 04.2 &2004/04 &9722 &2006/05        & 9.67  &85.0  & 5.2 &15.4&\nodata & 0         &0      &0       &0	     &F814W \\
MACSJ0717$+$3745 	&0.546 	&07 17 32.93 	&$+$37 45 05.4 &2004/04 &9722 &2005/10        & 8.67  &83.4  & 5.7 &16.9&2006/10 & 10.5  & 72.2 &6.2    &18.5   &F814W \\
MS0016.5$+$1654$^a$  &0.546 	&00 18 32.80 	&$+$16 26 06.9 &2002/01 &9292 &2006/06        & 6.65  &56.4  &1.3 &3.7   &\nodata & 0         &0      &0       &0	      &F775W \\
MACSJ0025$-$1222  	&0.584 	&00 25 30.23 	&$-$12 22 43.0 &2004/10 &9722 &\nodata         & 0        &0       &0      &0     &\nodata & 0        &0      &0       &0	      &F814W \\
MACSJ2129$-$0741  	&0.589 	&21 29 26.30 	&$-$07 41 26.2 &2003/09 &9722 &2005/06       & 8.64   &70.1  &1.7   &5.1   &\nodata & 0        &0       &0      &0	      &F814W \\
MACSJ0647$+$7015 	&0.591 	&06 47 49.78 	&$+$70 14 56.4 &2004/12 &9722 &2006/02       & 8.41   &71.6  &3.9   &10.3 &2006/11 & 8.54  & 74.6 &3.9 &10.4     &F814W \\
SDSSJ1004$+$4112      	&0.680 	&10 04 34.72 	&$+$41 12 45.0 &2004/04 &9744 &2005/12$^d$& 4.65   &59.0  &2.3  &12.6 &2007/01 & 9.68  &58.6 &2.3 &12.6      &F814W \\
MACSJ0744$+$3927 	&0.697 	&07 44 52.58 	&$+$39 27 26.7 &2004/02 &9722 &2005/12        & 8.43   &58.2 &4.1   &11.1&2006/12 & 8.97   &55.6 &4.0 &10.9      &F814W \\
MS1054.4$-$0321    	&0.833 	&10 57 00.20 	&$-$03 37 27.0 &2002/12 &9290 &2006/01        & 9.11   &43.3 &3.8   &11.3 &2007/01 & 10.4  &43.5 &3.9 &11.6       &F775W \\
CL0152$-$1357        	&0.835	&01 52 43.00 	&$-$13 57 20.0 &2002/11 &9290 &2005/06        & 10.0   &32.4 &2.7   &8.0  &2006/09 & 7.20   &34.8 &2.5 &7.5         &F775W \\
CLJ1226.9$+$3332  	&0.888 	&12 26 58.21 	&$+$33 32 49.4 &2003/04 &9033 &2006/01        & 10.7  &35.8  &1.1   &3.6  &2007/01 & 10.5   &37.8 &1.7 &5.5         &F814W 
\enddata
\tablenotetext{a}{MS0016.5$+$1654 and MS0451.6$-$0305 are also MACS
  clusters, MACSJ0018.5$+$1626 and MACSJ0454.1$-$0300, respectively.}
\tablenotetext{b}{References for cluster redshifts are Ebeling et
  al. (2007) for the MACS clusters, Ebeling et al. (2001) for
  CL0152$-$1357 and CLJ1226.9$+$3332, Tran et al. (1999) for
  MS1054.4$-$0321, and Oguri et al. (2004) for SDSS1004$+$41.}
\tablenotetext{c}{$\Delta t$, $L_B$, and $M$ of Epoch I are the same as those of Epoch II.}
\tablenotetext{d}{GO-10509.}
\tablenotetext{e}{Overlapping area between epochs [arcmin$^2$]. The imaging area of ACS is 10.5 arcmin$^2$.}
\tablenotetext{f}{Visibility time [days].}
\tablenotetext{g}{Stellar luminosity within the search area [$10^{12}\, {\rm L}_{B,\sun}$] (see \S~\ref{s.luminosity2}).}
\tablenotetext{h}{Stellar mass within the search area [$10^{12}\, {\rm M}_{\sun}$] (see \S~\ref{s.results}).}

\end{deluxetable*}


\begin{turnpage}
\tabletypesize{\scriptsize}
\begin{deluxetable*}{lllll|lllll|l}
\tablecolumns{10} 
\tablewidth{0 pt} 
\tablecaption{SN candidates$^a$ \label{table.cands}}
\tablehead{\multicolumn{4}{c}{Source}               &
           \colhead{Discovery}                                    &
           \multicolumn{5}{|c|}{Host}                           &
	    \colhead{Classification}                             \\
           \colhead{Name}                   &
           \colhead{RA (h~m~s)}                               &
           \colhead{Dec ($^\circ$~$'$~$''$)}                                     &
           \colhead{mag$^b$}                   &
           \colhead{epoch}                                    &
           \colhead{RA (h~m~s)}                                          &        
           \colhead{Dec ($^\circ$~$'$~$''$)}                         &
           \colhead{mag} &
           \colhead{Morphology}                                          &               
           \colhead{Spec $z$ (Ref)$^c$}                                          }          
\startdata
MACSJ0257-2325-cand1 & 02:57:08.479 & -23:24:24.38 & 25.4 & Jan 2004 & 02:57:08.507 & -23:24:24.55 & 22.9 &   Irr             & 0.3294 (O7) & Foreground\\
MACSJ0257-2325-cand2 & 02:57:12.425 & -23:27:05.08 & 26.8 & Jan 2004 & 02:57:12.425 & -23:27:05.08 & $\gtrsim27$& unresolved & N/A & Possible cluster SN Ia\\
MACSJ0257-2325-cand3 & 02:57:10.005 & -23:27:14.52 & 25.8 & Aug 2005 & 02:57:09.983 & -23:27:13.54 & 22.3 &   spiral       & 0.73 (FF1)     & Background\\
MACSJ0257-2325-cand4 & 02:57:11.476 & -23:27:19.72 & 26.5 & Jan 2004 & 02:57:11.464 & -23:27:19.45 & 23.6 &   E               &  \nodata       & Possible cluster SN Ia\\
MACSJ0257-2325-cand5 & 02:57:06.904 & -23:27:46.79 & 26.3 & Jan 2004 & 02:57:06.947 & -23:27:46.44 & 20.9 &    spiral       & 0.4039 (O7)& Foreground\\
\tableline
MACSJ0647+7015-cand1 & 06:47:30.004 & +70:14:54.53 & 25.8 & Feb 2006 & 06:47:29.79  & +70:14:54.12 & 22.5 & spiral/Irr & 0.619 (O8)   & Background\\
MACSJ0647+7015-cand2 & 06:47:38.030 & +70:16:16.72 & 26.8 & Dec 2004 & 06:47:38.024 & +70:16:16.05 & 22.2 & spiral     & 0.495 (O8)   & Foreground\\
MACSJ0647+7015-cand3 & 06:47:49.824 & +70:15:31.19 & 24.6 & Nov 2006 & 06:47:49.901 & +70:15:30.40 & 18.6 & spiral     & 0.365 (O8)   & Foreground\\
MACSJ0647+7015-cand4 & 06:47:59.661 & +70:15:19.51 & 25.2 & Nov 2006 & 06:47:59.626 & +70:15:19.37 & 23.7 & unclear   &   \nodata     & Possible cluster SN Ia\\
\tableline
MACSJ0717+3745-cand1 & 07:17:38.90   & +37:45:20.8  & 26.3 & Apr 2004 & 07:17:38.855  & +37:45:20.07 & 21.7 & spiral    & 0.55 (FF2)        & Cluster, CC \\ 
MACSJ0717+3745-cand2 & 07:17:31.444 & +37:44:36.12 & 24.4 & Oct 2006 & 07:17:31.519  & +37:44:37.58 & 20.0 &    E        & 0.4915 (MS)     & Foreground\\
MACSJ0717+3745-cand3 & 07:17:41.444 & +37:44:10.54 & 24.3 & Oct 2006 & 07:17:41.444  & +37:44:10.54 & 20.6 &    E        & 0.538 (MS)       & Cluster SN Ia\\
MACSJ0717+3745-cand4 & 07:17:40.548 & +37:45:06.60 & 23.3 & Oct 2006 & 07:17:40.548  & +37:45:06.60 & 21.6 & compact    &  2.084 (FSP)  &Background AGN \\
\tableline 
MACSJ0911+1746-cand2 & 09:11:16.407 & +17:47:40.04 & 24.9 & Oct 2005 & 09:11:16.407 & +17:47:40.04 & 21.6 &    S0        & 0.88 (E)           & Background AGN  \\
MACSJ0911+1746-cand4 & 09:11:16.959 & +17:46:48.42 & 25.1 & Dec 2006 & \nodata      & \nodata      &\nodata& No host        & \nodata           & Hostless, possible cluster SN Ia \\
MACSJ0911+1746-cand5 & 09:11:19.290 & +17:46:09.27 & 25.2 & Dec 2006 & 09:11:19.269 & +17:46:10.07 & 21.0 & Spiral     & 0.842 (FF4)    & Background \\
\tableline
MACSJ1149+2223-cand1 & 11:49:33.20  & +22:24:29.90 & 23.9 & Apr 2004 & 11:49:33.147 & +22:24:30.35 & 20.0     & E                  & 0.553 (FF5)& Cluster SN Ia \\
MACSJ1149+2223-cand2 & 11:49:35.734 & +22:22:18.34 & 24.5 & May 2006 & 11:49:35.726 & +22:22:18.41 & 23.6  &    E              &   \nodata          &Possible cluster SN Ia\\
\tableline
MACSJ2129-0741-cand1 & 21:29:28.356 & -07:41:34.44  & 25.0 & Sep 2003 & 21:29:28.321 & -7:41:34.77  & 23.1     &  Irr               & 0.87 (GY)  &  Background   \\
MACSJ2129-0741-cand2 & 21:29:24.993 & -07:42:22.76  & 24.4 & Sep 2003 & \nodata           & \nodata         & \nodata& No host     & \nodata      &  Hostless, possible cluster SN Ia \\
\tableline
MACSJ2214-1359-cand1 & 22:14:59.353 & -13:58:15.28 & 26.0 & Aug 2005 & 22:14:59.382 & -13:58:14.92  & 20.9 & Spiral        & 0.582 (FF1)        &  Background  \\
MACSJ2214-1359-cand2 & 22:14:57.012 & -14:00:11.49 & 23.7 & Oct 2003  & 22:14:57.012 & -14:00:11.49  & 19.6 &     E            & 0.503 (GYS)    & Cluster SN Ia\\
\tableline
CLJ1226.9+3332-cand1 & 12:26:54.552 & +33:33:56.38 & 25.8 & Jan 2006 & 12:26:54.508 & +33:33:56.17 & 20.5 & Spiral     & 0.59 (GYS) & Foreground\\
CLJ1226.9+3332-cand2 & 12:26:55.503 & +33:32:12.42 & 23.7 & Apr 2003 & 12:26:55.588 & +33:32:12.78 & 20.2 &   S0       & 0.9009 (E) & Possible cluster SN Ia \\
\tableline
MS1054.4-0321-cand1  & 10:56:56.262 & -03:37:51.04 &24.9 ($i$)& Dec 2002 & 10:56:56.287& -03:37:51.59 &19.8 (i)& Spiral  & 0.230 (FF5)& Foreground \\
MS1054.4-0321-cand2  & 10:56:57.862 & -03:37:47.77 &23.5 ($i$)& Dec 2002 & 10:56:57.862& -03:37:47.77 &21.3 (i)&  S0       & 0.8335 (VD)& Cluster SN Ia\\
\tableline
MS0451.6-0305-cand1  & 04:54:09.968 & -03:00:28.18 & 25.5 & Jul 2005 & 04:54:10.054  & -03:00:27.76 & 18.3 & Spiral     & 0.16 (FF3a) & Foreground \\
\tableline
CL0152-1357-cand1    & 01:52:43.099 & -13:55:19.76 &24.3 ($i$)& Jun 2005 & 01:52:43.099& -13:55:19.76 &23.2 (i) & compact & 1.27 (GY)  & Background AGN \\
CL0152-1357-cand2    & 01:52:37.99  & -13:56:25.69 &25.2 ($i$)& Jun 2005 & 01:52:38.06 & -13:56:25.5  &21.5 (i)& Spiral  & 1.12 (FS)  & Background \\
CL0152-1357-cand3    & 01:52:46.217 & -13:58:03.93 &24.1 ($i$)& Sep 2006 & 01:52:46.217& -13:58:03.93 &23.9 (i)& unclear & 0.839 (FS) & Cluster SN Ia \\
\tableline
SDSSJ1004+4112-cand1 & 10:04:33.086 & +41:12:31.20 & 25.2 & Dec 2005 & 10:04:33.075  & +41:12:30.34 & 22.3   &   spiral      & 0.753 (FF3b)& Background \\
SDSSJ1004+4112-cand2 & 10:04:30.601 & +41:14:10.66 & 25.9 & Dec 2004 & 10:04:30.645  & +41:14:10.67 & \nodata  &   \nodata     &    \nodata  (FF3b)       & Possible cluster SN Ia\\
SDSSJ1004+4112-cand3 & 10:04:31.007 & +41:14:13.59 & 26.3 & Jan 2007 & 10:04:31.047  & +41:14:13.10 & 21.6   &    E     & 0.75 (O7)  & Background \\
\enddata
\tablenotetext{a}{Coordinates are J2000.}
\tablenotetext{b}{Magnitudes are observed $I$ band, unless indicated.}
\tablenotetext{c}{Redshifts are based on the following observations: 
(FF1) Keck/LRIS       $1''$    longslit, Filippenko and Foley, 2005 Dec 3;
(FF2) Keck/DEIMOS $0.9''$ longslit, Filippenko and Foley, 2005 Dec 1; 
(FF3) Keck/DEIMOS (a) $0.9''$ longslit, (b) multislit, Filippenko and Foley, 2005 Dec 31; 
(FF4) Keck/LRIS       $1''$    longslit, Filippenko and Foley, 2006 Dec 20;
(FF5) Keck/DEIMOS multislit, Filippenko and Foley, 2007 Feb 16;
(FF6) Keck/LRIS       multislit, Filippenko and Foley, 2007 Jan 12;
(FS)   Keck/LRIS $1''$ longslit, Filippenko and Silverman, 2007 Nov 12;
(FSP) Keck/LRIS $1''$ longslit, Filippenko, Silverman, and Poznanski, 2008 Apr 26.
(GY)  Keck/LRIS,       $1''$    longslit, Gal-Yam, 2005 Aug 1;
(VD)  Keck/LRIS,       $1\farcs2$    longslit, Van Dokkum at al. (2000);
(E)     Keck/DEIMOS,  multislit, Ebeling, 2006 Jan;
(GYS) Keck/LRIS multislit, Gal-Yam and Sharon, 2007 Jul 16-17;
(MS)  Keck/LRIS, $1''$ longslit, M. Sullivan, 2006 Nov 22; 
(O7)  Keck/LRIS multislit, Ofek, 2007 Jan 22;
(O8)  Keck/LRIS multislit, Ofek, 2008 Jan 4. 
}
\end{deluxetable*}
\end{turnpage}

\begin{deluxetable*}{lllllp{1.5cm}lllll}
\tablewidth{0pt} 
\tablecaption{Subaru Imaging Data \label{tab.subaru}}
\tabletypesize{\scriptsize}
\tablecolumns{11} 
\tablehead{ 
\colhead{Cluster Name$^a$ }                                    &
           \multicolumn{5}{c}{UT Date Observed  [mm/yy]}     &
	  \multicolumn{5}{c}{Exposure time  [s]}\\ 
	  \colhead{}&
	  \colhead{$B$}&
	  \colhead{$V$}&
	  \colhead{$R$}&
	  \colhead{$I$}&
	  \multicolumn{1}{l}{$~~z'$}&
	 \colhead{$B$}&
	  \colhead{$V$}&
	  \colhead{$R$}&
	  \colhead{$I$}&
	  \colhead{$z'$}
	  }
\startdata
MACSJ0257         & 09/05& 12/02&  12/00& 12/07& 12/02 &  1440&  2160&  5280& 2400& 2700 \\
MACSJ0454$^b$ & 11/05& 11/05&  03/05& 12/01& 12/06 & 1440 &  2160 &  3240 & 2160 & 1440 \\
MACSJ0647 & 11/05 & 09/03 & 02/04 & 02/04 & 02/04 & 1440 & 2160 & 2880 & 2880 & 2160 \\
MACSJ0717 & 02/04 & 12/02 & 12/00 & 12/00 & 12/02 & 1440 & 2160 & 2880 & 1440 & 1620 \\
MACSJ0744 & 02/04 & 04/03 & 12/02 & 04/03 & 04/03 & 1440 & 1440 & 2880 & 3240 & 2160 \\
MACSJ0911 & 11/05 & 04/03 & 04/03 & 04/03 & 04/03 & 2880 & 2160 & 2880 & 1200 & 1620 \\
MACSJ1149  & 12/06 & 04/03 & 04/03 & 12/00 & 04/03 & 1440 & 2160 & 2880 & 1200 & 1620 \\
MACSJ1423  & 07/03 & 06/02 & 06/02 & 06/02 & 06/02 & 1920 & 2160 & 2400 & 2160 & 1440 \\
MACSJ2129  & 07/03 & 06/02 & 06/01 & 06/01 & 06/02 & 2880 & 1440 & 2880 & 2880 & 1440 \\
MACSJ2214  & 11/05 & 09/03 & 07/03 & 07/04 & 07/04 & 1440 & 2160 & 2880 & 2160 & 1620 \\
CL1226     & 05/06 & 06/02 & 12/00 & 12/00 & 04/03 & 2160 & 2160 & 2880 & 1920 & 1080 \\
\enddata
\tablenotetext{a}{Full cluster names are listed in Table~\ref{table.clusters2}.}
\tablenotetext{a}{MACSJ0454.1$-$0300 is also named MS0451.6$-$0305.}
\end{deluxetable*}


\begin{deluxetable*}{llllll}
\tablecolumns{6} 
\tablewidth{0 pt} 
\tablecaption{Cluster-member SN candidates \label{table.clustercands}}
\tablehead{\colhead{Name}                   &
           \colhead{$M_V^a$}                                &
           \colhead{$M_B^a$}                                &
           \colhead{$R^b$}                                 &
           \colhead{Classification}                            &
           \colhead{Note}                              }          
\startdata
MACSJ0257-2325-cand2 &   $-$14.7  & \nodata  & 0.48    & possible  & Unknown $z$\\
MACSJ0257-2325-cand4 &    $-$15.0  & \nodata & 0.51    &  possible & Unknown $z$, early red-sequence host$^c$ \\
MACSJ0647+7015-cand4 &    $-$16.6  & \nodata & 0.37    &  possible & Unknown $z$\\
MACSJ0717+3745-cand1 &    $-$15.3  & \nodata & 0.56    &  not SN Ia & Cluster $z$, likely CC~SN \\ 
MACSJ0717+3745-cand3 &    $-$17.4  & \nodata & 0.69    & likely      & Cluster $z$, early host\\
MACSJ0911+1746-cand4 &    $-$16.4  & \nodata & 0.49    &  possible & Hostless\\
MACSJ1149+2223-cand1 &    $-$17.8  & \nodata & 0.32    &  likely      &  Cluster $z$, early host\\
MACSJ1149+2223-cand2 &    $-$17.2  & \nodata & 0.60    & possible & Unknown $z$, early host\\
MACSJ2129-0741-cand2 &    $-$17.5  & \nodata & 0.38    & possible & Hostless\\
MACSJ2214-1359-cand2 &    $-$17.8  & \nodata & 0.02    &  likely     &  Cluster $z$,  early host\\
CLJ1226.9+3332-cand2  & \nodata      & $-$18.6 &0.38     & likely     &  Cluster $z$, early host\\
MS1054.4-0321-cand2   & \nodata      & $-$18.9 &0.26     & likely     &  Cluster $z$,  early host\\
CL0152-1357-cand3       & \nodata     & $-$18.4  & 0.44    &  likely     &  Cluster $z$,  brightness suggests SN Ia\\
SDSSJ1004+4112-cand2 &    $-$15.8  & \nodata & 0.67     &  possible & One of the possible hosts is at cluster $z$\\
\enddata
\tablenotetext{a}{Absolute magnitude at the time of detection, K-corrected to rest-frame $V$ for candidates at $z<0.7$ and to 
rest-frame $B$ for candidates at $z>0.7$. In  cases of unknown redshift, the SN candidate is assumed to be at the
  cluster redshift.}
\tablenotetext{b}{Projected distance from brightest cluster galaxy, in
  Mpc.}
\tablenotetext{c}{The host has the same $i-z'$ color as the cluster
  red-sequence galaxies; see Figure~\ref{fig.colormag1}.}
\end{deluxetable*}

\begin{figure}[]
\rotatebox{0}{\scalebox{1.15}{\plotone {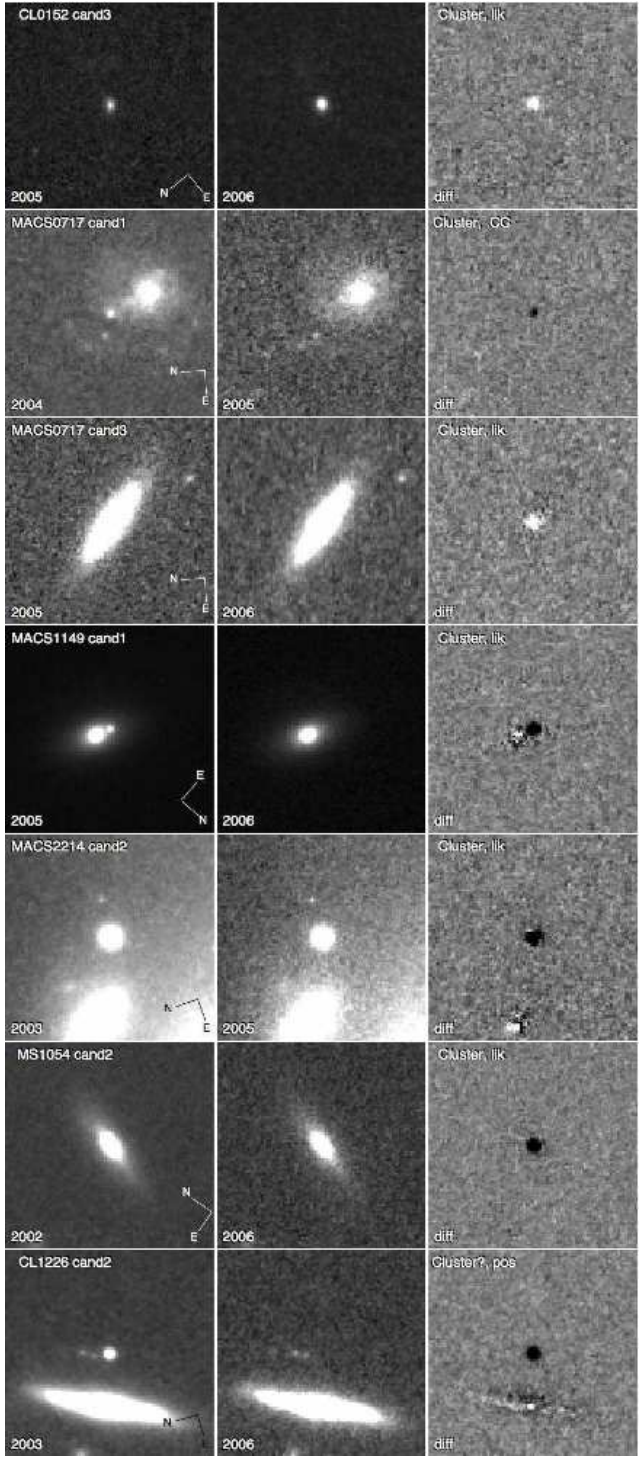}}}
\caption{Discovery images of the possible and likely cluster-member SN
  candidates.  In each row of three thumbnails, we show the first and
  second epochs in the left and middle frames, respectively. In the
  right panel we show the difference, as a subtraction of the earlier
  epoch from the later epoch. The frames are each $4\farcs4$ wide, and
  centered on the transient candidate.  
}

\label{fig.cluster_tiles}
\end{figure}
\addtocounter{figure}{-1}

\begin{figure}[]
\rotatebox{0}{\scalebox{1.15}{\plotone {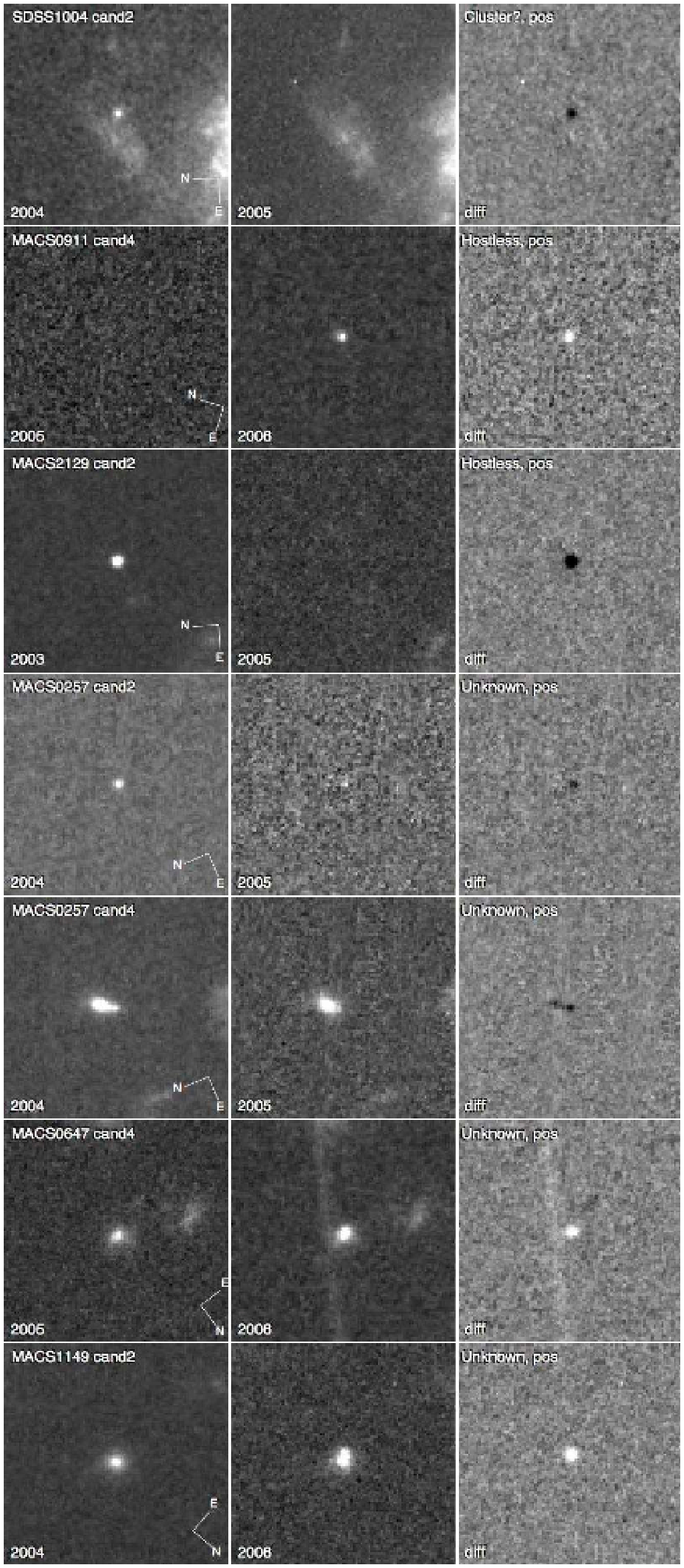}}}
\caption{Continued.}
\end{figure}

\begin{figure}[]
\rotatebox{0}{\scalebox{1.15}{\plotone {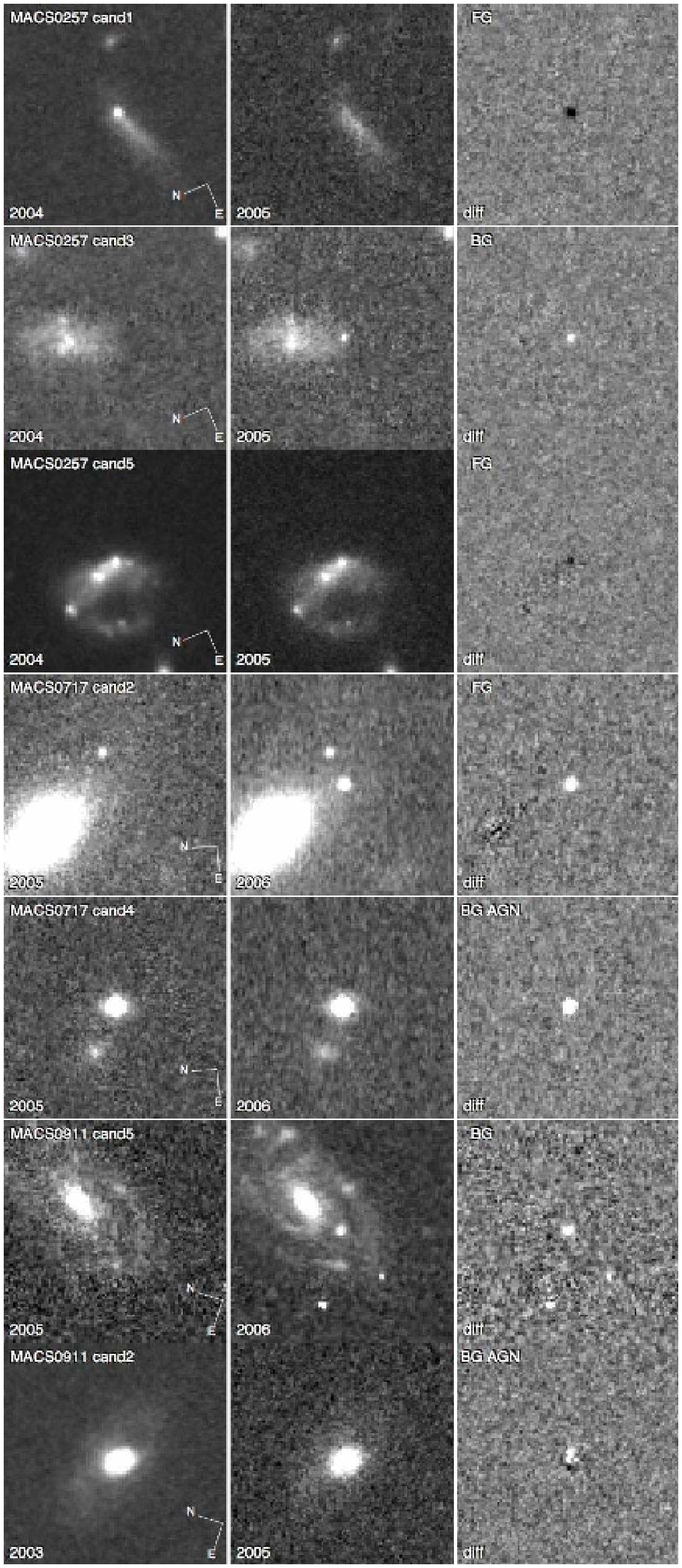}}}
\caption{Same as Figure \ref{fig.cluster_tiles}, for non-cluster candidates.}
\label{fig.field_tiles1}
\end{figure}

\addtocounter{figure}{-1}

\begin{figure}[]
\rotatebox{0}{\scalebox{1.15}{\plotone {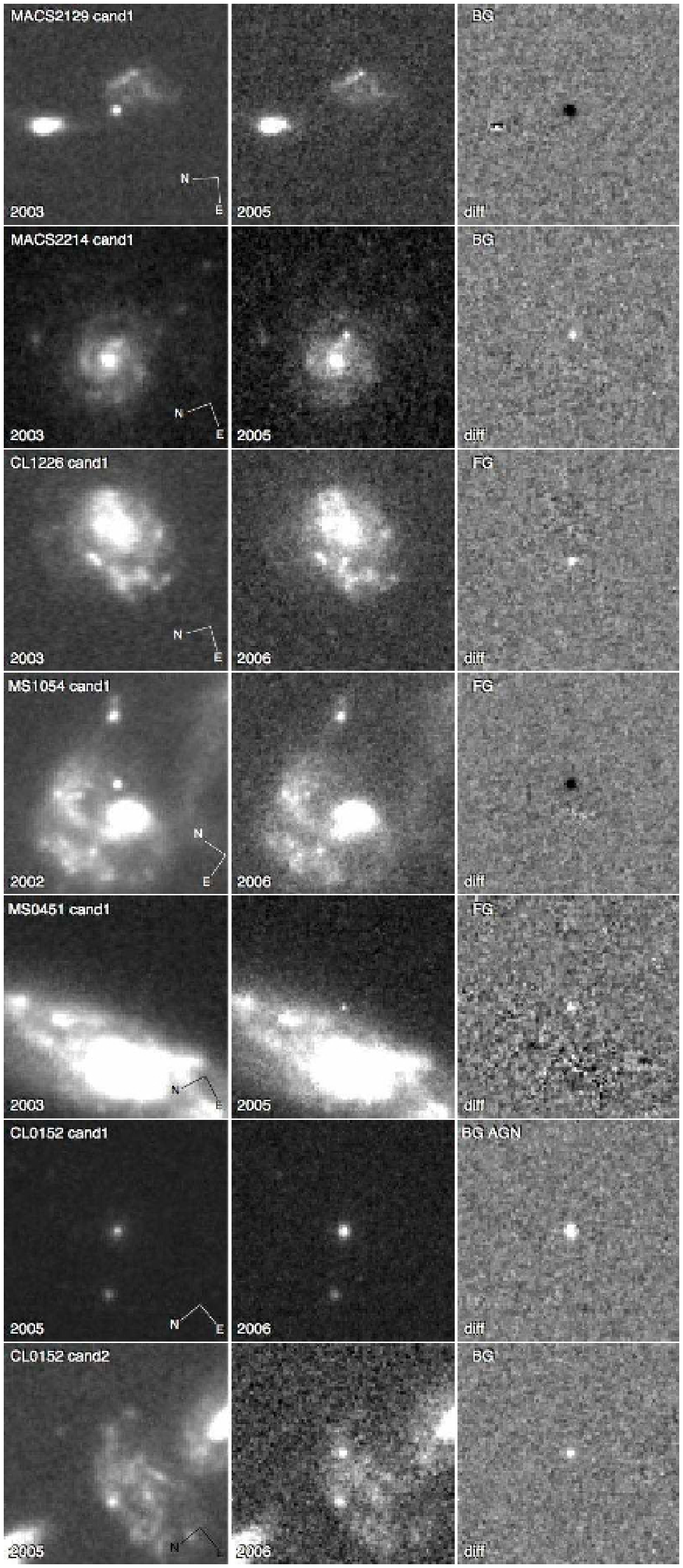}}}
\caption{Continued. }
 \label{fig.field_tiles2}
\end{figure}
\addtocounter{figure}{-1}

\begin{figure}[]
\rotatebox{0}{\scalebox{1.15}{\plotone {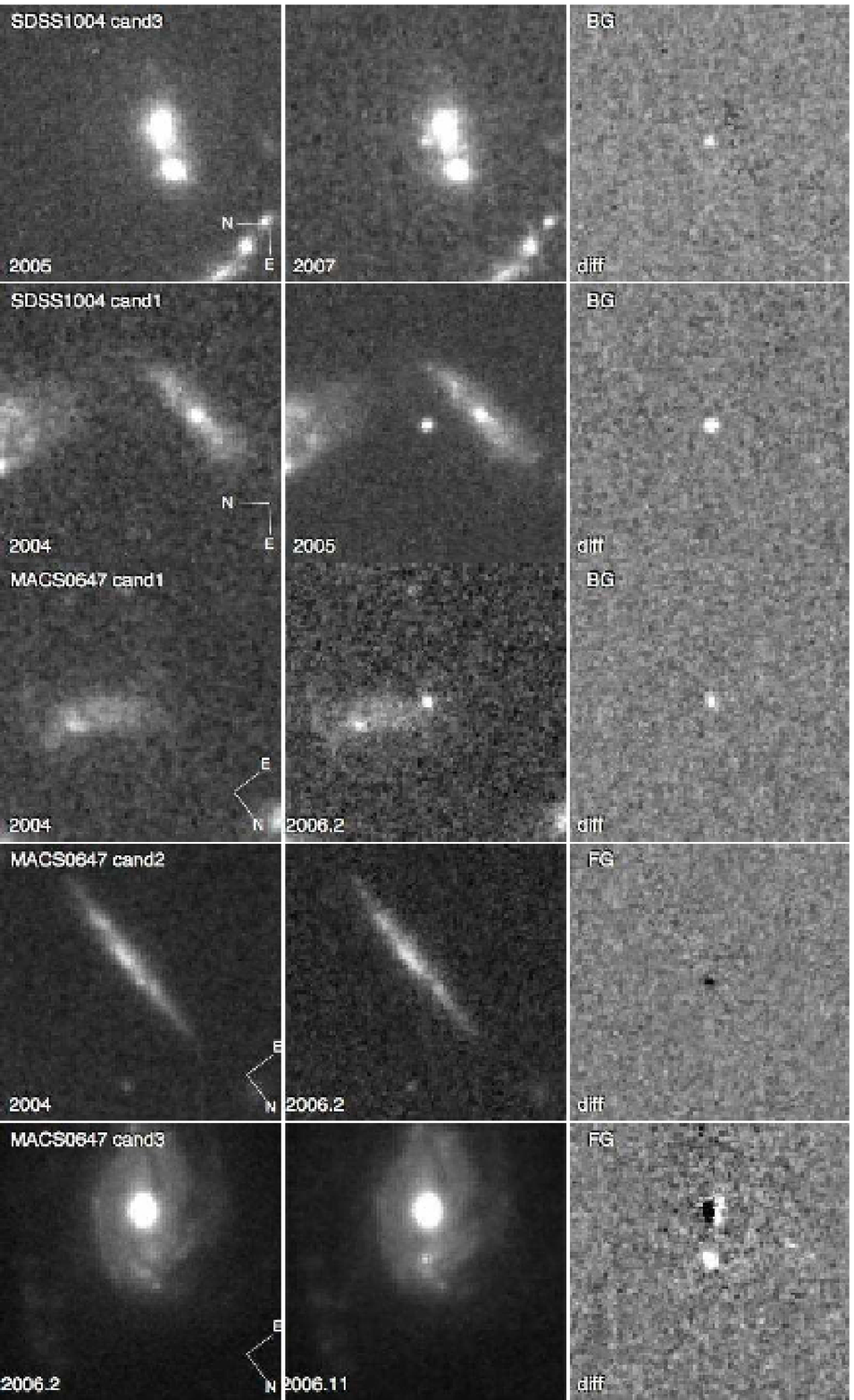}}}
\caption{Continued.}
\label{fig.field_tiles6}
\end{figure}

\section{Subaru Images}\label{s.subaru}

Our data analysis and derivation of SN rates made use of auxiliary
ground-based data (see \S~\ref{s.calcrate},
below).  Eleven of the clusters in our survey were imaged as part of a
study of MACS clusters (Ebeling et al. 2007) using the SuprimeCam
wide-field imager mounted on the Subaru 8.2-m telescope at Mauna Kea
(Miyazaki et al. 2002), covering a field of view of
$\sim 34' \times 27'$ per cluster. Table~\ref{tab.subaru} lists the
clusters for which we have obtained Subaru data, the observation
dates, and the exposure times. 
The 2.5$\sigma$ limiting magnitude is typically $R=25.7$ mag (AB).
Details regarding the Subaru data,
including data reduction and photometry, can be found in Kartaltepe et
al. (2008).  Briefly, the observations took place between 2000 and
2007 in variable conditions, with seeing ranging from $0\farcs6$ to
$1\farcs1$.  The data were reduced with the standard SuprimeCam
pipeline (Donovan 2007).  Photometric zeropoints were derived from
overlap with the SDSS, or (for clusters outside the SDSS footprint)
using 3~s exposures of nearby SDSS fields interlaced between the
cluster observations. 
The zeropoint uncertainty is $\sim 0.1$~mag, and results from the
different Subaru and SDSS filter passbands, and from the non-simultaneity
of the interlaced calibration photometry.

Object catalogues for each cluster were created using SExtractor
(Version 2.4.3; Bertin \& Arnouts 1996) in ``dual-image'' mode (i.e.,
detecting objects in one image, while performing photometry on
another) with the $R$-band image as the reference detection image.
Star/galaxy separation was based on the SExtractor parameter
\texttt{MU\_MAX} (peak surface brightness above the background
level). Since the light distribution of a source (e.g., its half-light
radius) scales with magnitude, stars and other point sources populate
a well-defined locus in a \texttt{ MU\_MAX}/\texttt{MAG\_AUTO} plane,
and can be excluded from the catalog (e.g., Bardeau et
al. 2005). In addition, objects with peaks sharper than the PSF are
not real astronomical objects, and can be flagged as artifacts.  These
catalogues were used as supplementary information for candidate
classification (\S~\ref{s.snalt}), and for measuring the cluster
stellar luminosities (\S~\ref{s.luminosity2}), below.


\section{Host-Galaxy Spectroscopy}\label{s.followup}

Nearly all of the SN candidates were too faint to be observed
spectroscopically, even at the time of discovery, and even more so
when pre-allocated ground-based follow-up observing time
arrived. Cluster membership was therefore established through
spectroscopy of the host galaxy.  In Table~\ref{table.cands}, we
summarize the spectroscopic information acquired for each of the
candidates.  Follow-up spectra were obtained primarily using the
10-m Keck telescopes in Hawaii, either with the Low Resolution
Imaging Spectrometer (LRIS; Oke et al. 1995) or the
Deep Imaging Multi-Object Spectrograph (DEIMOS; Faber et al. 2003),
in longslit or multislit configuration.  The target
lists were selected from the \textit{HST} image, and supplemented by
other interesting objects in the field, such as gravitationally lensed
galaxies and cluster members.  For DEIMOS multislit spectroscopy, we
filled the field of view with additional targets drawn from wide-field
imaging with Subaru or SDSS.  Table~\ref{table.cands} lists the
dates, instrument, and observers of each spectroscopic observation.

Multislit observations were reduced and analyzed as follows. After
standard bias and flatfield corrections, we combined all the
observations of the same field into a deep, cosmic-ray-cleaned
two-dimensional spectrum. We compared each multislit spectrum with
known night-sky lines, and calibrated the wavelength range using the
IRAF\footnote{IRAF (Image Reduction and Analysis Facility) is
  distributed by the National Optical Astronomy Observatories, which
  are operated by AURA, Inc., under cooperative agreement with the
  National Science Foundation.} tasks \texttt{IDENTIFY},
\texttt{FITCOORDS}, and \texttt{TRANSFORM}. Finally, we used the IRAF
task \texttt{APALL} to subtract the background, trace the continuum of
the object, and extract the one-dimensional spectrum from the
calibrated image\footnote{Some of the spectra were reduced using tools 
developed in the MATLAB environment (Ofek et al. 2006).}.  
The resulting spectrum was rebinned and searched
for common galaxy emission and absorption lines.  Longslit data were
reduced in a similar manner, and in addition were also flux calibrated
using spectra of standard stars from the same observing nights. This
process allows, in principle, a comparison of the spectral shape of
the object with those of template spectra, and derivation of a
redshift even in the absence of emission or absorption features.  
In practice, the continuum
signal from the host galaxies was generally weak, and the redshift
determination is based on emission or absorption lines.


\section{Candidate Classification}\label{s.snalt}

As our transient candidates are not spectroscopically confirmed, we
must consider the possibility that some or all of them are not SNe.
The survey discussed in this paper is similar to the one reported by
Gal-Yam et al. (2002), which was based on archival \textit{HST} images
of galaxy clusters, obtained with the Wide Field Planetary Camera 2
(WFPC2). While our survey is superior in resolution to the WFPC2
survey, and has a more uniform observation scheme, both surveys reach
a similar depth, the search methods and efficiency simulations are
done in the same manner, and the time span between epochs is similar.
As discussed by Gal-Yam et al. (2002), transient events that could, in
principle, mimic SNe include solar-system objects, variable stars in
our Galaxy or in other galaxies, AGNs, or gamma-ray burst (GRB)
afterglows. As Gal-Yam et al. (2002) argue, most of these transients
cannot be confused with SNe. First, all but two of the candidates are
clearly associated with galaxies. The probability for a chance
association is small, $<2-6\%$ (see \S~\ref{s.survey}).  The proper
motions of asteroids or Kuiper-belt objects would have been detected
in our long-exposure images. With the exposure span within one 
{\it HST} orbit (typically $\sim45$ min), we can detect proper motions
greater than $\sim 0 \farcs 015$ hr$^{-1}$.  At the dates and
coordinates of our observations, the parallax of a Kuiper-belt object
at 50 AU due to the Earth's motion is more than 
$0\farcs 4$,
well above our detection
limit.  Variable stars in other galaxies are too faint to mimic SNe.


Variable stars in our own Galaxy, to be undetected in one of the
epochs, would have to be distant.  For example, an M5 flare star 
with absolute magnitude $M_I=9$~mag in its quiescent state
(Allen 1973) would have to be at a distance $> 40$ kpc 
to be fainter than our $I=27$~mag detection limit. 
The local density of halo stars with
$0.09<M/{\rm M}_\odot<0.71$ is of order $10^{-4}~{\rm pc}^{-3}$ (Gould et
al. 1998).  A number of recent studies (see, e.g., Cignoni et
al. 2007, and references therein) find an approximately $r^{-3}$
profile for the outer halo, and thus the stellar density at 40 kpc is $10^{-4}
(8.5~{\rm kpc}/40~{\rm kpc})^{3} \approx 10^{-6}~{\rm pc}^{-3}$.
Integrating this density from 40 kpc to, say, 80 kpc gives a surface
density of $\sim 10^4$ deg$^{-2}$.  There could thus be several
hundred outer-halo late-type stars in the 15 ACS fields. Based on
this, the two hostless candidates we have found could, in principle,
be distant Galactic M stars that had flared into visibility on one
epoch.  However, the flare would have had to be by $\gtrsim 2-2.5$~mag 
to bring the object from below the detection limit,
$I=27$ mag, to the
observed brightnesses of the two hostless transients, $I=25.1$
and $I=24.4$~mag, respectively. Although stellar
flares can be bright in the ultraviolet, and reach amplitudes of
$\Delta U \approx 5$ mag, their amplitudes in red optical bands are much
smaller (e.g., Eason et al. 1992; Allred et al. 2006; Zhilyaev et
a. 2007), generally $\Delta I \lesssim 1$~mag.  Furthermore, a recent
SDSS-based study by Kowalski et al. (2009) of M-dwarf flaring
frequency and magnitude finds that the flaring duty cycle is strongly
correlated with Galactic height. Beyond
300 pc above the disk, only $10^{-5}$ of their individual observations, which
have cadences of several days,
catch an M star
in its flaring phase. Thus, apart from the implausibility of a
large-amplitude $I$-band flare, there is only a $\sim 10^{-3}$
probability to start with, that any of the $\sim 100$ outer-halo 
M stars in the ACS fields would
be caught flaring in our observations.
It is therefore highly unlikely that the hostless
candidates are optical flares of Galactic stars.


For candidates that are in the centers of their associated hosts, an
AGN nature would have been revealed by our spectroscopy. Since GRBs
are often associated with some CC~SNe (Galama et al. 1998;
Hjorth et al. 2003; Stanek et al. 2003; Malesani et al. 2004; Pian et
al. 2006; see Woosley \& Bloom 2006 for a review), we argue that in
addition to being unlikely, they would have affected only our
classification as SNe~Ia or CC~SNe, not the identification as SNe.  
We thus conclude that essentially
all of our transient candidates are bona fide SNe.


\section{Classification of Cluster SN Candidates}\label{s.clustercands}

Cluster membership of a SN candidate was decided according to the SN
host-galaxy redshift and the cluster velocity dispersion (Ebeling et
al.  2007). We classified SN candidate hosts as cluster members if
their redshift indicated that their velocity is within $2\sigma$ of
the cluster recession velocity, where $\sigma$ is the velocity
dispersion of the cluster.  We note that the velocity dispersion of
galaxies in clusters as massive as those in our sample is relatively
high; the mean value for MACS clusters in our sample is $\sim 1300$ km
s$^{-1}$ (Ebeling et al. 2007).

\begin{figure}[h]
\rotatebox{0}{\scalebox{1.2}{\plotone {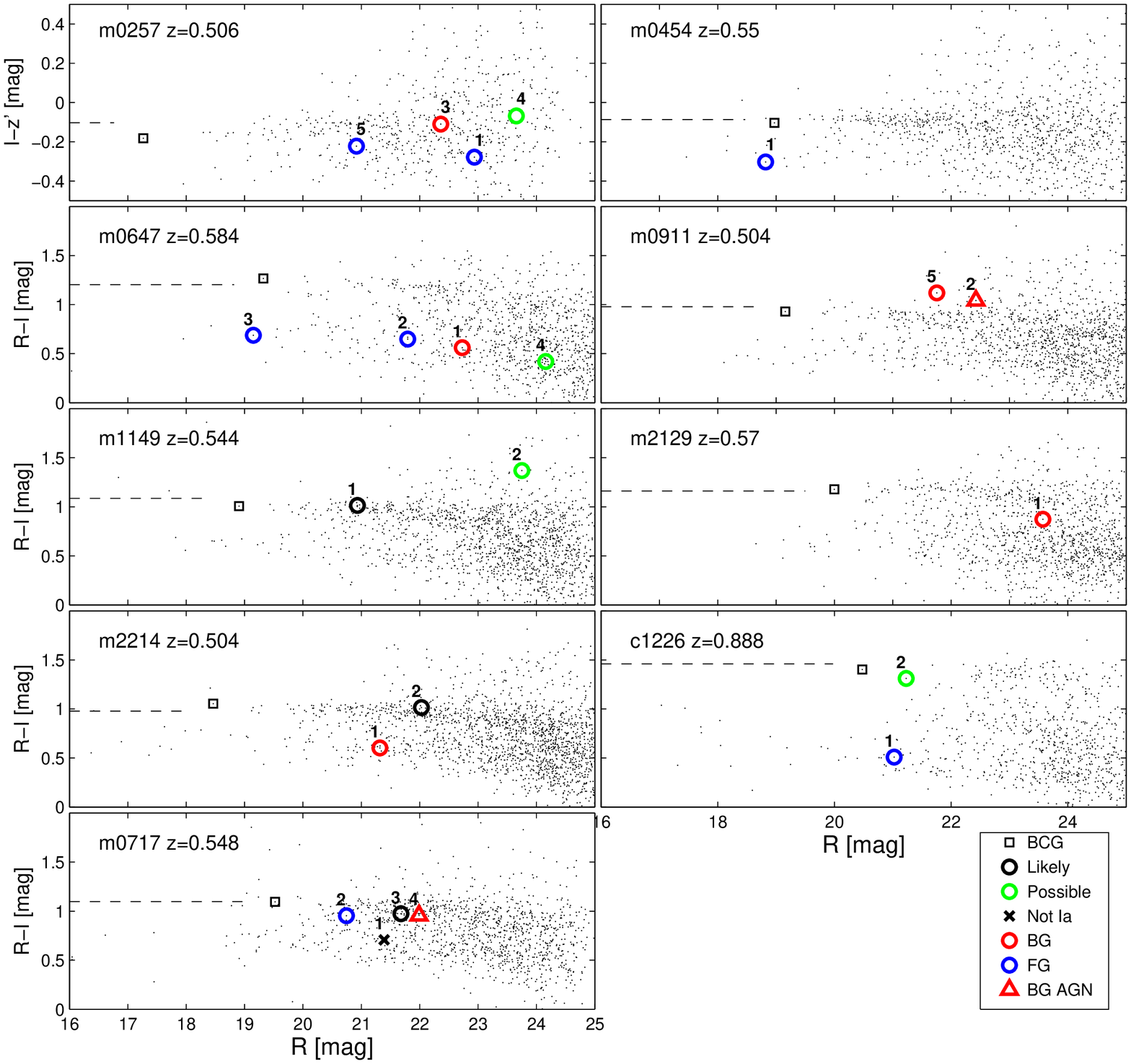}}}
\caption{Color-magnitude diagrams for the galaxies in the cluster
  fields with SN candidates for which we have Subaru data. In each
  plot, the brightest cluster galaxy is marked with a black
  square. A dashed line indicates the color of an elliptical galaxy at the 
  cluster redshift. Host galaxies are marked with circles (see legend).  }
 \label{fig.colormag1}
\end{figure}

\begin{figure}[h]
\rotatebox{0}{\scalebox{1.2}{\plotone {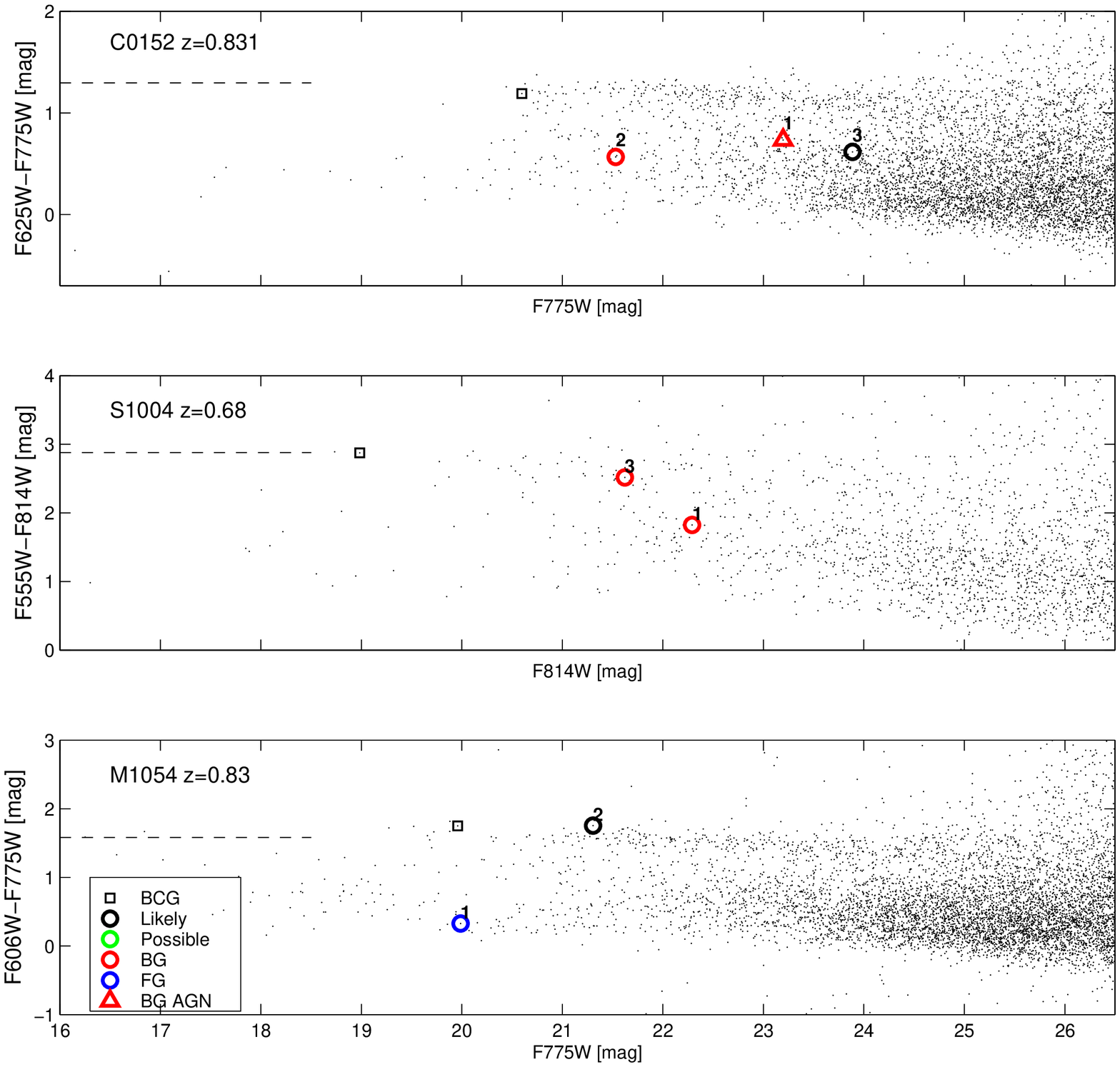}}}
\caption{Same as Figure 4, but for cluster fields for which we
  do not have Subaru data.  The magnitudes are measured in the
  available ACS bands in the archive.}
 \label{fig.colormag2}
\end{figure}

Next, although we have no spectroscopic classification, some clues
about the type of the SN candidates exist.  Since CC~SNe are
exceedingly rare in non-star-forming environments (e.g., Hakobyan et
al. 2008), we classify all non-AGN candidates in cluster early-type
galaxies as SNe~Ia. The high resolution of ACS allows classification
of resolved host galaxies through morphology, which can be supported
by color information where available. In cases where the morphological
type is unclear, we compare the location of the host in
color-magnitude space with that of the cluster red sequence
(Figures~\ref{fig.colormag1} and \ref{fig.colormag2}). For the MACS
clusters, we use photometric catalogs of multiband Subaru images (see
\S~\ref{s.subaru} for details). Multiband archival \textit{HST}/ACS
data are also available for some of the clusters.  The classification
of each individual candidate is described below.\\

\subsection{Candidates with Cluster-Member Hosts}

\paragraph {Likely cluster SNe~Ia.} 
Based on ACS morphology, colors, and Subaru photometry, four of the
cluster candidate hosts are undoubtedly early-type galaxies: 
MACS0717 cand3, MACS1149 cand1, MACS2214 cand2, and MS1054
cand2. In the three cases where the candidate is not clearly separated
from the galaxy core, we confirmed via spectroscopy that it is not an 
AGN.  We therefore classify these events as likely cluster SNe~Ia.

\paragraph {CL0152 cand3, likely cluster SN~Ia.}
This candidate is detected in the second epoch with $I \approx 24.1$~mag,
slightly offset from the center of a cluster galaxy. Although the
galaxy is bluer than the cluster red sequence, at $z=0.8391$ the SN
absolute magnitude, $M_B \approx -18.4$, is strongly suggestive of a SN~Ia,
given the scarcity of overluminous CC~SNe (less than 10\% of
 CC SNe at maximum brightness are as bright in a volume-limited sample; 
Li et al. 2010, in preparation).

\paragraph {MACS0717 cand1.}
The SN candidate appears in the archival epoch, in a spiral galaxy at
the cluster redshift, and was observed in two filters, $V$ and $I$.
The observed $V-I$ color ($\sim1.6$ mag) and magnitude ($I \approx
26.3$, corresponding to $M_V \approx -15.3$ mag at $z=0.548$) are
consistent with a CC~SN.  Young SNe Ia would be too bright (expected
to be $<23$ mag), and older SNe Ia would be too red.  We therefore
argue that this candidate is not a SN~Ia.

\subsection{Candidates with Ambiguous Hosts}

CL1226 cand2 appears in the archival epoch, $1\farcs12$ from the core of an S0
galaxy at the cluster redshift. The $V-I$ color ($\sim2.1$ mag) and magnitude
($I \approx 23.7$ mag) are consistent with a SN~Ia, 9--10 days after maximum 
brightness. However, the candidate is also positioned close to a blue emission
knot that is possibly a part of a disrupted spiral galaxy at unknown redshift, 
centered $\sim5''$  NNW of the cluster member.
For SDSS1004 cand2, the results of the spectroscopy are ambiguous,
with indications that the SN may be in a cluster member galaxy.
However, it is also on the outskirts of a foreground spiral
galaxy at $z=0.27$.
Adopting a conservative approach, we therefore classify these
candidates as ``Possible'' cluster SNe~Ia.

\subsection{Hostless Candidates}

Two of the SN candidates, MACS0911 cand4 and MACS2129 cand2, have no
detectable host at the limiting magnitude of the coadded ACS
images. If these are indeed cluster events, they occurred 30~kpc and
20~kpc away from any galaxy, respectively. The absolute magnitudes of
these candidates, $M_V=-16.4$ and $M_V=-17.5$, are consistent with SNe~Ia
at these clusters redshifts. Their projected distances from the 
brightest cluster galaxies (BCGs) are
0.49 Mpc and 0.38 Mpc, respectively, at the clusters redshifts.  The
$V-I$ color of MACS0911 cand4 does not rule out any SN type at the
cluster redshift.  These, if intergalactic cluster events (Gal-Yam et
al.  2002), are most probably SNe Ia, since intergalactic CC SNe
should be very rare; there is little or no star formation in the
intergalactic medium (Gal-Yam et al.  2003), 
and even if a CC SN progenitor is ejected from
its host galaxy, it cannot move far before exploding since it is
short lived.  The progenitors of SNe~Ia, on the other hand, have ample
time to reach 30--70 kpc from their host galaxy before explosion.

\subsection{Candidates Without Measured Host Redshift}

We do not have secure redshift measurements for four of the SN
candidate hosts.  Spectroscopy was attempted (unless indicated), but
the objects proved to be too faint and/or without emission lines.  The
host of MACS0257 cand2 is too faint for spectroscopy, with
$I\gtrsim27$~mag.  The luminosity of the SN candidate does not rule it
out as a SN~Ia.  MACS0257 cand4 is offset from the core of a resolved
host with an elliptical morphology.  The host colors are consistent
with the cluster red sequence in the available Subaru bands ($I$,
$z'$).  MACS1149 cand2 is offset from the core of a host with an
early-type morphology. The host colors are consistent with the red
sequence in $V-I$, but are redder than cluster galaxies in $R-I$ and
$R-z'$. Spectroscopy of this galaxy failed due to a mistake in mask
designs, as a result of which the galaxy spectrum fell off the CCD.
MACS0647 cand4 is offset from the core of a
resolved host with unclear morphological type. 
Subaru colors indicate that it is $\sim1.5$ mag bluer
than the cluster red sequence.

\subsection{Classification Summary}
We conclude that five of the cluster events are likely SNe~Ia, and one
is likely a CC~SN.  To these we add, as possible cluster SNe~Ia, the
two hostless SNe, the two SNe with ambiguous hosts,   
and the four candidates for which we do not have
measured redshifts.  In terms of radial distribution, the cluster SNe
are found at projected distances of up to 0.7 Mpc from the BCG (see
Table \ref{table.clustercands}).

In order to reduce our uncertainties due to misclassification, 
when calculating the SN~Ia rate we impose 
a magnitude cut and remove from our sample candidates fainter than 
26 mag. This changes the SN count to five likely and six possible 
cluster SNe~Ia.


\section{SN Rate Calculation}\label{s.calcrate}

With the sample of potential cluster SNe~Ia constructed above, we can
now derive a SN rate for our cluster sample. The SN rate per unit
stellar luminosity is calculated as follows:
\begin{equation}\label{eq.SNR2} 
{ R}_{\rm Ia} = \frac{N}{\sum\limits_{j}{\Delta t_j L_{band,j}}},
\end{equation}
where $N$ is the number of SNe, $\Delta t_j$ is the visibility time
(or ``control time,'' the time during which a cluster SN~Ia is
above the detection limit of the $j$th image), $L_{band, j}$ is the
cluster luminosity within the search area of the $j$th image in the 
chosen photometric
band, and the summation is over all the survey images. The details of
each element in the calculation are given below.
To account for the statistical nature of some of the quantities that enter 
the rate calculation, we conduct a Monte-Carlo simulation 
in which we measure the rate many
times, each time assigning values to the measured quantities by
drawing them from a distribution (see \S~\ref{s.errors}). The final rate and its 
uncertainty is measured from a histogram of the results. 

\subsection{Visibility Time}

The visibility time depends on the detection efficiency, the peak
luminosity, and the shape of the light curve at a given redshift
and filter. In principle, it can also depend on the time interval
between observations, as a SN is less likely to be detected via
subtraction if it has a similar brightness at both epochs. In
practice, since all of our comparisons are for epochs that are separated
by at least one year, old SNe would have had enough time to decline
below our detection limit, and thus the visibility time is not
affected.  We calculate the effective visibility time from
\begin{equation}\label{eq_deltat2}
\Delta t_j = \int^{\infty}_{-\infty} {\eta[m(t)]dt},
\end{equation}
where $m(t)$ is the SN~Ia light curve in the image bandpass (either
F814W or F775W) at the given redshift, and $\eta[m(t)]$ is the
detection probability as a function of SN magnitude.  We describe
below each step in this calculation.


\subsubsection{Detection Efficiency Estimate}\label{S.efficiency2}

\begin{figure}
\rotatebox{0}{\scalebox{1.2}{\plotone {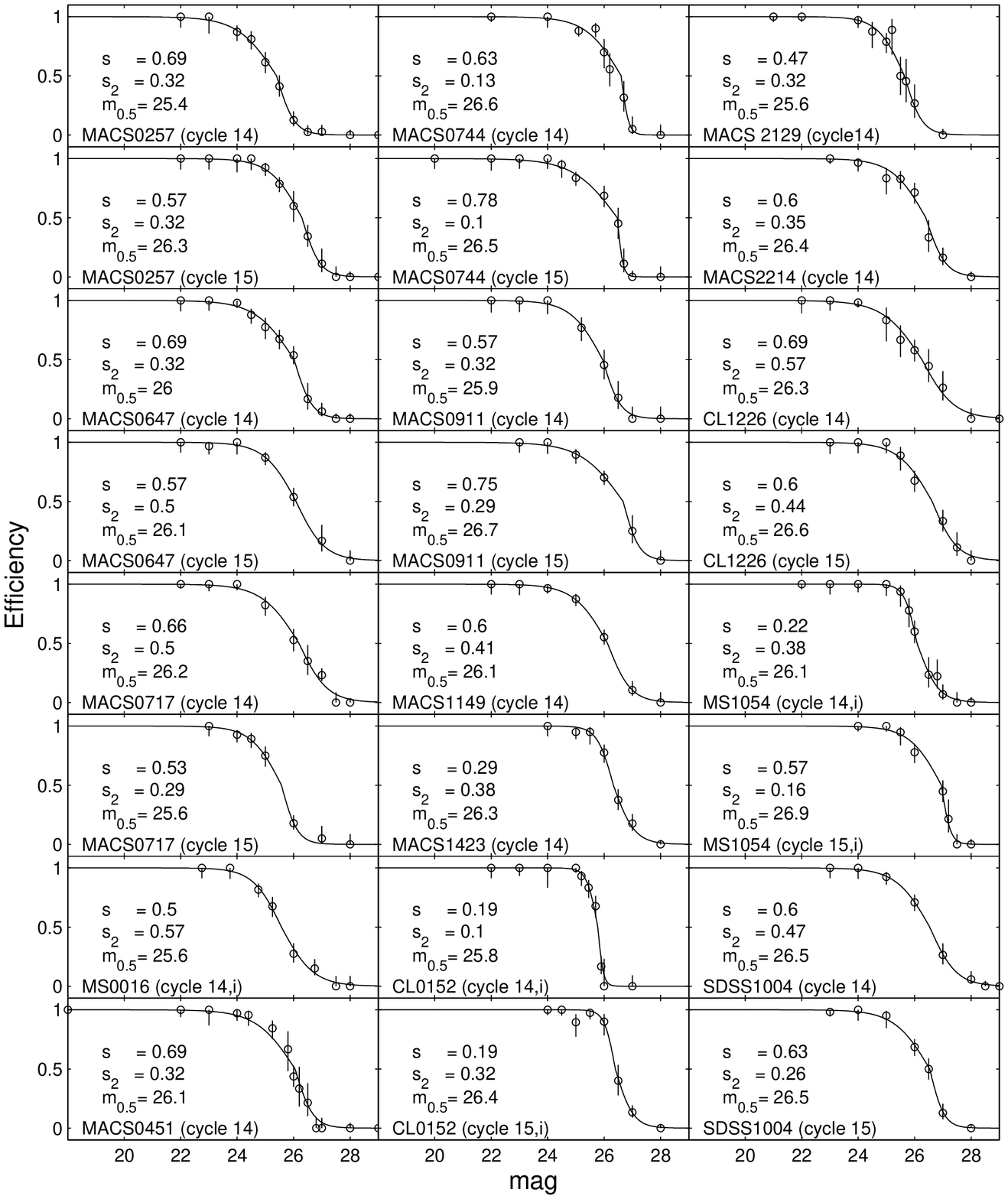}}}
\caption{Detection efficiency curves for a point source in a difference 
  image, as a function of magnitude. The difference images are Epoch II
  vs. Epoch I (indicated by ``cycle 14'')  and Epoch III vs. Epoch II 
  (indicated by ``cycle 15''). Magnitudes are measured in the difference 
  image, and are Vega-based $I$-band magnitudes, except where 
  indicated by ``$i$.''  
  Cluster names are as marked.  Circles mark the
  fraction of detected fake SNe, with error bars based on a binomial
  distribution. Solid lines are the best-fit efficiency curves with
  the indicated parameter values, parametrized as in
  Eq.~\ref{eq.effhst}.}\label{fig.HSTeff}
\end{figure}

To determine the detection efficiency of our survey, we conducted
efficiency simulations following the scheme detailed by Gal-Yam et
al. (2002) and Sharon et al. (2007). To each field, we added some 200
fake SNe, in a range of magnitudes, and with a spatial distribution
that follows the flux of the galaxies. The simulated images then
underwent the same search procedure as the real data, and the number
of SNe that were recovered in each magnitude bin was noted.  We find
that, while the efficiency function strongly depends on the limiting
magnitude of the image, its shape is also sensitive to other
attributes of the field and of the observation. Although \textit{HST}
images do not suffer from atmospheric distortion of the PSF, the cores
of bright galaxies are not perfectly subtracted in difference images,
resulting in residuals that can hide faint transients. This effect is
amplified when two epochs are not obtained at the same position angle,
due to position-dependent variations in the PSF.  We experimented with
several PSF-matching techniques (e.g., Gal-Yam et al. 2008) and
determined that a simple subtraction is sufficient for our purposes.
We find that there is 100\% efficiency in all epochs to
detect SNe at magnitudes brighter than 23, even in the cores of bright
galaxies. The efficiency differs from field to field, and even between
epochs of the same field.  Figure~\ref{fig.HSTeff} shows the results
of our efficiency simulations. We parametrized our efficiency curves
as a function of magnitude $m$ with the function
\begin{equation}\label{eq.effhst}
\eta(m;m_{0.5},s,s_2) = \left\{ \begin{array}{ll}
      \left(1 + e^{\frac{m-m_{0.5}}{s}}\right)^{-1}, & \mbox{$m\le m_{0.5}$}\\
      \left(1 + e^{\frac{m-m_{0.5}}{s_2}}\right)^{-1}, & \mbox{$m>m_{0.5}$},\\
      \end{array} \right.
\end{equation}
where $m_{0.5}$, $s$, and $s_2$ are free parameters that are fit to
the simulated efficiencies: $m_{0.5}$ is the magnitude at which the 
efficiency drops to $0.5$, and $s$ and $s_2$ determine the range of 
$m$ over which $\eta$ changes from $1$ to $0.5$ and from $0.5$ to 
zero, respectively.  

In the final rate calculation, we apply a magnitude cut at 26 mag, 
by setting the efficiency above 26 mag to zero, and rejecting from the sample
SNe that were detected above that value.

\subsubsection{Light Curves}\label{S.lightcurves}

The light curves of SNe~Ia exhibit an empirical inhomogeneity, where
luminous SNe tend to rise and decline more slowly than
subluminous ones.  
This means that luminous SNe will be above the
detection limit for longer than subluminous ones, resulting in an 
overall visibility time that is correlated with SN brightness.
The correlation
between peak magnitude and light-curve shape can be parametrized by a
stretch relation (Phillips 1993; Perlmutter et al. 1999; see
Leibundgut 2001 for a review), which exhibits an intrinsic root-mean
square (rms) scatter of $\sim 0.2$ mag. In this paper, we use the form
$M_s = M_{s=1} - \alpha(s-1)$, $t_s = t_{s=1}\times s$, as described
by Perlmutter et al. (1999), where $s$ is the stretch factor ($s=1$
means an unstretched light curve), $M_s$ is the absolute $B$-band
magnitude, and $\alpha = 1.47$, based on the empirical results of Knop
et al. (2003).  The peak-magnitude distribution of SNe~Ia also depends
on the host-galaxy properties.  By comparing light curves of SNe in
different environments, Sullivan et al. (2006b) have shown that SNe~Ia
in elliptical galaxies tend to be dimmer (with a smaller stretch
factor) than those in star-forming galaxies (Hamuy et al. 2000;
Howell et al. 2001).
 
For each cluster, we compile a set of light curves relevant to the
cluster redshift and imaging band for several light-curve shapes and
peak magnitudes.  We start with a rest-frame, non-stretched, $B$-band
template light curve from Nugent et al. (2002) and transform it to
various stretched light curves, using the stretch relation above, and
the measured distribution of elliptical host-galaxy stretch factors
from Sullivan et al. (2006b).  For consistency, we use the same
non-stretched peak $B$-band absolute magnitude as Sullivan et
al. (2006a), $M_B = -19.25$ mag (for $H_0 = 70$ km s$^{-1}$
Mpc$^{-1}$).  We assume an uncertainty of $0.15$ mag in $M_B$, from
the dispersion in peak magnitudes of local SN~Ia light curves after
applying a stretch correction (Guy et al. 2005). This uncertainty is
taken into account in our error budget (see \S~\ref{s.errors}).

A set of multi-epoch spectral templates from Nugent et al. (2002) are
normalized to fit the $B$-band rest-frame stretched light curve, and
redshifted according to the cluster redshift. We then combine the
redshifted spectra with the \textit{HST} bandpass in which the cluster was
observed, to obtain a light curve for each combination of stretch,
redshift, and bandpass in our survey.
When calculating the visibility time for a particular image, we draw
for each SN in the Monte-Carlo simulation a stretch factor (with its
corresponding properly normalized light curve) from the Sullivan et
al. (2006b) distribution of stretch factors.
Table~\ref{table.clusters2} lists the visibility time for each epoch, for
the most likely stretch factor. We note that since epochs I and II are 
searched against each other, the values enumerated in 
Table~\ref{table.clusters2} (i.e., visibility time, $L$, $M$, and the 
search area) are the same for epochs I and II and are only listed once.


\subsection{Cluster Stellar Luminosity}\label{s.luminosity2}

A SN rate in a targeted galaxy population needs to be normalized by
the stellar luminosity or mass of that population. These stellar
properties, in turn, must be measured to some level of accuracy.  Our
rate measurement accuracy is limited by the small number of SNe that
were discovered, with a $1\sigma$ statistical error of order 30--40\% (lower
and upper Poisson errors for 10 events; see \S~\ref{s.staterrors},
below). An accuracy of 10\% in the cluster stellar luminosities will 
therefore be more than satisfactory for our purposes. We generally follow 
the route detailed by Sharon et al. (2007), and perform ``aperture
photometry'' of the cluster light in several bands, as follows.  For
the MACS clusters and CLJ1226.9+3332, we use the Subaru data centered
on each cluster (see \S~\ref{s.subaru}). For most of the MACS
clusters, we were able to use calibrated $V$, $R$, and $I$ images,
with additional $B$- and $z'$-band imaging for some.

Using the galaxy catalogs described in \S~\ref{s.subaru}, 
we measure the total flux in galaxies within a given aperture, and
subtract from it the flux per unit area in galaxies in a ``background'' area at
the outskirts of the images, multiplied by the area of the
aperture. This way, we statistically subtract the contribution of
``background'' objects (those in front of and behind the cluster). The
aperture in which we measure the cluster stellar luminosity is the
exact search area (i.e., the area of overlap between epochs).  The
``background'' is sampled in an annulus with area of at least $100$
arcmin$^2$, with an inner radius of at least $7'$ from the
cluster center. These ``background'' annulus radii vary among the
sample clusters, depending on the angular size of the cluster and the
available Subaru field of view, and take into account masking of
bright foreground objects.  The completeness of the Subaru data varies
between clusters and depends on exposure time and observational
conditions.  For each cluster, we estimate the completeness magnitude as
the turnover magnitude in its galaxy-magnitude histogram in the
$R$ band.
We ignore in our calculation galaxies fainter than the completeness
magnitude, and correct the total luminosity accordingly as explained
below. To avoid contamination from bright low-redshift objects, we
also ignore galaxies that are brighter than the BCG.

The luminosities of the four clusters for which we do not have Subaru
data were measured in the same manner, but from the {\it HST} data. Since
the ACS field of view is too small to select ``background'' annuli
around the clusters, we assumed a universal background, drawn from the
archival {\it HST}/ACS imaging of The Great Observatories Origins Deep
Survey (GOODS; Giavalisco et al. 2004).

For comparison with other SN rate measurements, we convert the
observed net integrated galaxy-light fluxes in the available bands to
rest-frame luminosities in several bands and form a cluster spectral
energy distribution (SED).  To each cluster SED, we fit redshifted
template spectra of combinations of several types of galaxies, using
synthetic photometry. For elliptical galaxies, we assume the Bruzual
\& Charlot (2003) synthesis of stellar population with a single
formation epoch at $z=3$. Other galaxy types are represented by
templates from Kinney et al. (1996).  Typically, within each search
area (i.e., the innermost $\sim1$ Mpc of each cluster in our survey),
the best fit was reached for a combination of 70--80\% elliptical-galaxy 
flux and 20--30\% Sbc-galaxy flux.  The fact that there is a
non-negligible fraction of blue galaxies is consistent with recent
measurements of the red fraction of galaxies in clusters in the
redshift range of our survey (Loh et al. 2008) due to the
Butcher-Oemler (1978, 1984) effect.  
The luminosities in the desired
rest-frame bands are measured via synthetic photometry of the best-fit
template combination, scaled to fit the observed net cluster flux.

To account for undetected cluster galaxies, we correct the luminosity
by multiplying it by the fraction of light that comes from the faint
end of a Schechter (1976) luminosity function
\begin{equation}
C = \frac{\int _0^\infty {L \Phi (L)dL} } { \int _{L_{\rm lim}(m_{\rm lim},z)}^\infty  L \Phi(L)dL},
\end{equation}
where $\Phi(L) dL = \Phi^*(L/L^*)^\alpha {\rm exp}(-L/L^*)d(L/L^*)$.
We adopt $\alpha= -1.00\pm0.06$ and $M^*=-22.01\pm0.11$ mag as the 
mean values for the
$g$-band luminosity function parameters in clusters (Goto et al. 2002).
We note that the depth of the Subaru images enables detection of
galaxies down to $M_I = -17$ mag, $\sim 5$ mag fainter than $M^*$,
and thus the necessary correction is small, typically 1--5\% for the
lower-redshift clusters, and up to $\sim11\%$ for the clusters at
$z>0.8$.

Finally, to account for passive evolution of the dominant elliptical 
galaxy component over the redshift range of the cluster sample, 
the Bruzual \&
Charlot (2003) component is passively evolved backward or forward,
as appropriate, to the mean visibility-weighted redshift of the sample,
$<\!\!z\!\!> = 0.6$ (see Table~\ref{table.clusters2}).  
The Sbc component is not evolved in this calculation. This
correction amounts to a 2\% change in the total luminosity.

\subsection{Error Budget}\label{s.errors}

In this section we estimate the sources of uncertainty, both
statistical and systematic.  While the statistical errors can be
propagated in a straightforward manner, the systematic errors affect
the final result in a more complicated way, and may be correlated. To
assess the overall systematic uncertainty, we calculate the SN rate by
performing a Monte-Carlo simulation in which we measure the rate many
times, each time assigning values to the measured quantities by
drawing them from a distribution centered on the best value, with a
width according to the uncertainty of this value. Where applicable, we
draw the values from a measured distribution and otherwise assume a
Gaussian distribution.

\subsubsection{Statistical Errors}\label{s.staterrors}

The counting of SN explosions obeys Poisson statistics, from which we
derive the statistical uncertainties. Contrary to the results
presented for $z \approx 0.2$ clusters by Sharon et al. (2007), the number
of secure cluster SNe in the present survey is not certain, due to
uncertainty in SN classification and redshift, and is between 5 and
11. We will consider this classification uncertainty as a systematic
error below. If we adopt for the central number of cluster SNe~Ia the
mean in this range, 8, the $1\sigma$ Poisson errors are $+49\%,
-35\%$ (Gehrels 1986).

\subsubsection{Systematic Errors} 

\paragraph {SN classification uncertainties. } 
A significant source of error in our rate derivation is the possible
misclassification of SNe.  Since our sample is not spectroscopically
confirmed, some of the SN candidates may possibly be CC SNe at the
cluster redshifts. CC SNe are preferentially located in star-forming
regions, and are rarely found in the elliptical galaxies that dominate
galaxy cluster environments.  Although some cluster star formation may
still be in progress, especially in high-redshift clusters, Saintonge
et al. (2008) find that such activity tends to be outside the central
1~Mpc, while all of our cluster candidate SNe are within 0.7~Mpc in
projection. Nevertheless, while many of the cluster-SN candidates in
our sample occurred in elliptical galaxies, we note that some are
associated with galaxies that have a late-type morphology. We will
therefore conservatively allow for a maximal misclassification error.
 
As described in detail in \S\ref{s.clustercands}, we set a firm lower
limit on the number of detected cluster SNe~Ia, from the five
candidates in early-type cluster galaxies. Eight candidates are
considered possible cluster SNe~Ia, of which two are fainter than 
26 mag and are rejected from our sample.
This sets the upper limit due to classification uncertainty at 11. 

\paragraph {Luminosity error.}
The derived cluster stellar luminosities depend on several
assumptions, such as the choice of area from which to draw the
background, photometric errors, and the assumption of the galaxy
templates with which the cluster SED was converted to rest-frame
luminosity.  To account for these uncertainties, we perform a
Monte-Carlo simulation, similar to those presented by Sharon et
al. (2007).  To assess the uncertainty introduced by our choice of
background area, we explored different choices of backgrounds for each
cluster. We find that for a $\lesssim 10\%$ variation in the inner
radius of the background annulus, the measured luminosity changes by
$\lesssim 5\%$.  Skewing our measurement of the fraction of elliptical
light in the cluster by 0.1 results in $\lesssim 10\%$ change in the
measured luminosity.  Assigning a $10\%$ uncertainty to the luminosity
of individual clusters in the Monte-Carlo simulation results in a
3.4\% rms variation of the SN rate.

\paragraph {Efficiency error.}
As explained in \S\ref{S.efficiency2}, we represent each efficiency
curve by a continuous function that we fit to the results of
efficiency simulations. To assess the errors induced by this process,
and by the fact that the number of fake SNe from which the efficiency
is derived is finite, we consider a range of efficiency-function
parameters distributed normally around the best-fit value, 
with a 10\% dispersion. In each
realization in the Monte-Carlo simulation we draw a set of parameters
from this distribution, resulting in an uncertainty due to the
efficiency estimation of $1.6\%$.

\paragraph {Visibility time error.}
The visibility time depends on the shapes and peak magnitudes of the
SN light curves at the cluster redshifts. These values are correlated
through the stretch relation (see \S\ref{S.lightcurves}).  We explore
a range of stretch parameters, following the distribution published by
Sullivan et al. (2006b). In each realization in the Monte-Carlo
simulation we draw a stretch factor from this distribution, and assign
a light curve accordingly. Since the visibility time is not linear
with a change of the stretch factor, the resulting distribution of SN
rates is no longer centered on the value that is calculated from the
most probable stretch factor. The inferred uncertainty due to the
visibility time error is $2.7\%$.

Accounting for all of the above systematic errors in the Monte-Carlo
simulation, we get an overall systematic uncertainty of $5\%$. This
uncertainty is, at present, negligible compared to the statistical
errors, but will become relevant in future surveys that detect more
than several hundreds of SNe.

\subsection{Results}\label{s.results}

The visibility-time-weighted mean redshift of our cluster sample is 
$<\!\!z\!\!> = 0.6$.  
For the adopted central value of $N_{\rm Ia}=8$, our measured SN rate 
per unit $B$-band stellar luminosity for this cluster sample is 
\snuB~SNu$_B$\, \snuBstat ~(statistical)\, \snuBclass ~(classification)\,
\snuBsys ~(systematic).

The SN rate per unit luminosity at $<\!\!z\!\!> = 0.6$ cannot be compared 
easily to rates at low redshifts, even in the absence of any star
formation, because of significant passive luminosity evolution.
For example, between $z=0.6$ and $z=0$, a passive population fades
by about a factor of 2 in blue bands (Bell et al. 2003; van der Wel et al.
2005). To facilitate such a comparison, we also list in Table~\ref{tab.rates}
a rate in units of SNu$_{B,0}$ which is a cluster SN rate at $<\!\!z\!\!> = 
0.6$ per unit stellar luminosity, 
but after passively evolving forward, to $z=0$, the elliptical
 Bruzual \& Charlot (2003) component fit to each cluster's photometry 
(the blue component, which contributes negligibly to the mass, is ignored 
in this calculation).

To derive the rate normalized by stellar mass, we follow Mannucci et
al. (2005), who converted $K$-band galaxy luminosities to mass using
the mass-to-light ratio derived by Bell et al. (2001) and observed
galaxy $B-K$ colors.  We derive the rest-frame $g-r$ colors of each cluster from
the best-fit combination of template spectra.  The stellar masses of the red and
blue components of each cluster
are then estimated from the color-dependent stellar mass-to-light ratio
derived by Bell et al. (2003), 
${\rm log}_{10}({\rm M}_{\sun}/{\rm L}_{g,\sun}) = -0.499 + 
1.519(g-r)$ (see Mannucci et al. 2005 for a discussion of the
validity of this ratio for our purpose). The total stellar mass in each search area
is enumerated in Table~\ref{table.clusters2}. The SN rate is calculated
from Eq. \ref{eq.SNR2}, in which we replace the stellar luminosity
with the inferred total mass within the search area of each cluster.  
For $N_{\rm Ia}=8$, the resulting SN~Ia rate per unit stellar mass is 
\snuM~SNu$_M$ \, \snuMstat ~(statistical)\, \snuMclass 
~(classification)\, \snuMsys ~(systematic). 
The different rates derived in this work are listed in Table~\ref{tab.rates}.

\begin{deluxetable}{lllll}
\tablewidth{0pt} 
\tablecaption{Cluster SN~Ia rate at $<\!\!z\!\!> = 0.6$\label{tab.rates}}
\tablecolumns{5} 
\tablehead{ 
 \colhead{Units$^a$}                  &
\colhead{SN Rate }                  &
	  \colhead{Statistical}&
	  \colhead{Classification}&
	  \colhead{Systematic}   \\
  \colhead{}                  &
\colhead{ }                  &
	  \colhead{Error}&
	  \colhead{Error}&
	  \colhead{Error}  
 }
\startdata
 SNu$_M$ & \snuM & \snuMstat &\snuMclass   & \snuMsys    \\
 SNu$_B$ &\snuB & \snuBstat &  \snuBclass   & \snuBsys  \\
 SNu$_{B,0}$ & \snuBO & \snuBOstat &  \snuBOclass   & \snuBOsys  \\
\enddata
\tablenotetext{a}{SNu$_B$ denotes SNe (100\, yr\, $10^{10}\, {\rm L}_{B,\sun})^{-1}$. 
 SNu$_M$ denotes SNe (100\, yr\, $10^{10}\, {\rm M}_{\sun}$)$^{-1}$], 
and ${\rm SNu_{B,0}} $ denotes a rate 
normalized to cluster luminosity that is passively evolved $z=0$.}
\end{deluxetable}

\section{Discussion and Summary}\label{s.discussion}

\begin{figure}
\rotatebox{0}{\scalebox{1.2}{\plotone {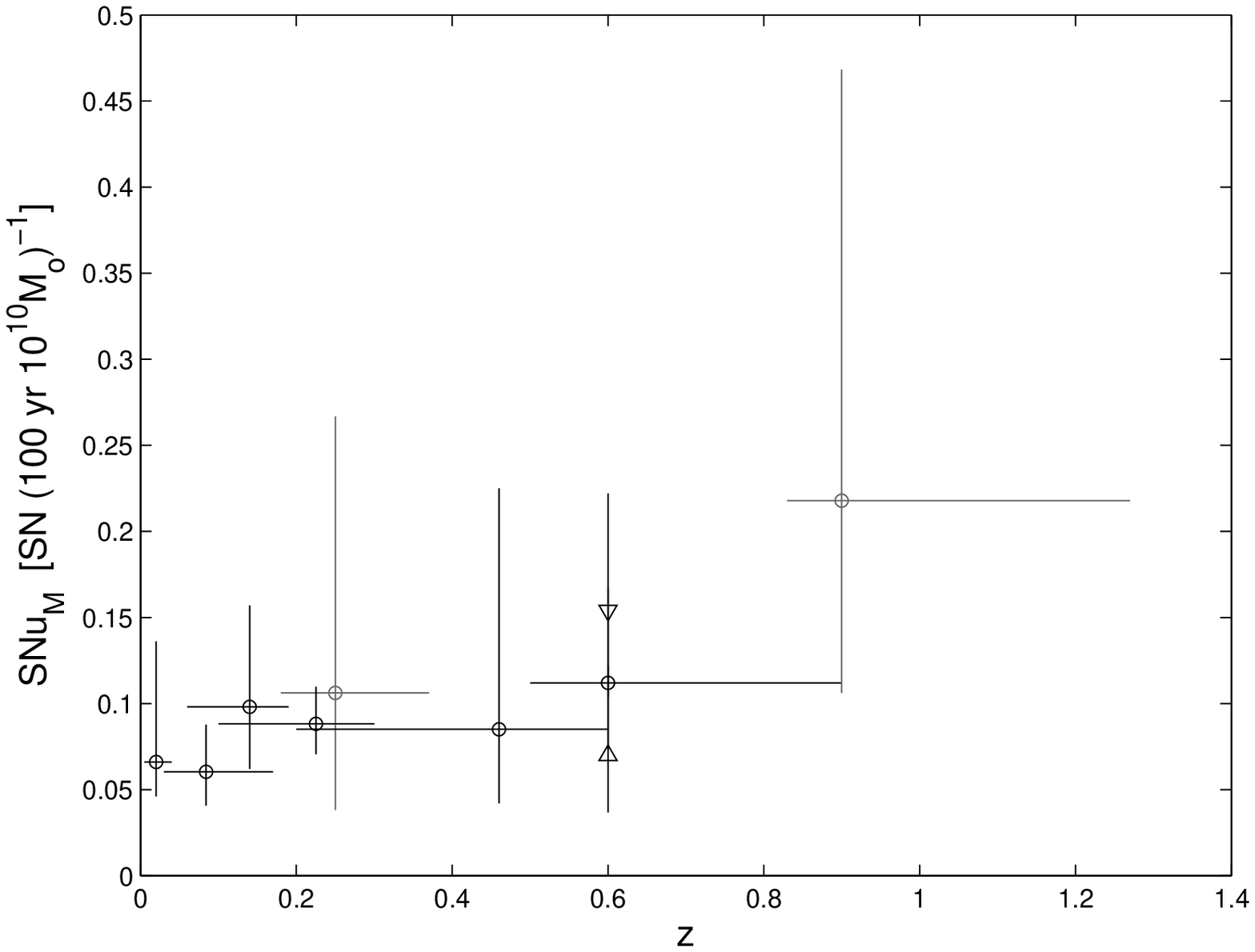}}}
\caption{Cluster SN rates from (in order of increasing redshift) 
  Mannucci et al. (2008), Dilday et al. (2010), Sharon et al. (2007), 
  Dilday et al. (2010), Gal-Yam et al. (2002),
  Graham et al. (2008), this work, and Gal-Yam et al. (2002). Triangles
  represent the SN rate derived from the upper and lower limits on the
  observed number of cluster SNe~Ia, 11 and 5, respectively. Vertical
  error bars are 1$\sigma$ uncertainties, and horizontal error bars
  show the cluster samples' redshift ranges. }
 \label{fig.ratez}
\end{figure}

Our newly measured SN rate at $z\approx 0.6$ is the most accurate to
date for clusters at such high redshifts. Combined with previous
cluster-rate measurements, we are, for the first time, in a position
of being able to examine the evolution of the SN rate in clusters over
the greater part of cosmic history.  Figure~\ref{fig.ratez} shows the
cluster SN rate as a function of redshift, at 
$<\!\!z\!\!> = 0.02$ (Mannucci et al. 2008), 
$<\!\!z\!\!> = 0.08$ and $0.22$ (Dilday et al. 2010),
$<\!\!z\!\!> = 0.15$ (Sharon et al. 2007),
$<\!\!z\!\!> = 0.25$ and $0.90$ (Gal-Yam et al. 2002),  
$<\!\!z\!\!> = 0.46$ (Graham et al. 2008), and 
$<\!\!z\!\!> = 0.6$ (this work). 
For consistency, the rates from
Gal-Yam et al. (2002) were converted from SNu$_B$ to SNu$_M$
using the ratio found by Sharon et al. (2007) for the lower redshift bin,
and using the ratio found in this work for the higher redshift bin.  
The SN rate in the redshift bin studied in this work is consistent with
the lower redshift rate measurements (to within uncertainties), and shows
that there is, at most, only a slight increase of cluster SN rate with
redshift.


In a companion paper (Maoz et al. 2010b), we analyze the
results and discuss the implications of the observed rates on SN
progenitor models, and on the role of SNe in the metal enrichment of
the ICM. 
However, some conclusions emerge directly from the low SN
rate observed out to $z\approx 1$.  Independent of any model, the
integral over the observed SN~Ia rate per unit mass between $z=1$ and
0 gives the total number of SNe per unit stellar mass.  If multiplied
further by the mean iron yield of a SN~Ia, $0.7\, {\rm M}_{\odot}$ (e.g.,
Mazzali et al. 2007), one obtains $M_{\rm Fe}/M_*$, the ratio of the
iron mass produced in clusters over that cosmic period to the
present-day stellar mass.  Assuming, for example, a constant SN~Ia
rate of 0.1 SNu$_{\rm M}$ over the past 6~Gyr, as implied by
Figure \ref{fig.ratez}, gives $M_{\rm Fe}/M_* \approx 0.0004$. In contrast,
the observed ratio in present-day clusters, after subtracting the
expected contribution from CC~SNe, is $[M_{\rm Fe}/M_*]_{\rm obs} \approx
0.004$ (see compilation and analysis of Maoz 2008), an order of
magnitude higher.  Thus, only a small fraction, $\lesssim 10\%$, of
the iron mass could have been produced by SNe-Ia between $z=1$ and 0. This 
conclusion is strengthened if one assumes, instead of $0.7\, {\rm M}_{\odot}$,
the lower iron yields associated with subluminous SNe~Ia. Such SNe tend 
to occur in the early-type galaxies that dominate galaxy cluster cores.
The factor $\sim 2$ increase in ICM iron abundance between these two
redshifts, recently reported by Balestra et al. (2007) and Maughan et
al. (2008), cannot be the effect of new iron production during this
time interval, and must instead be a redistribution
effect. Alternatively, the abundance evolution may not be real (Ehlert
\& Ulmer 2009), in which case the non-evolution is fully consistent
with our SN results.

Our measurements constrain the fraction of intergalactic SNe Ia
(Gal-Yam et al. 2003). Assuming the two hostless SNe are intergalactic
cluster events, and that only the 6 most secure events in our sample
are indeed SNe Ia in cluster galaxies, we get an upper limit on the
relative fraction of intergalactic events of 2/8. This is very similar
to the fraction measured by Gal-Yam et al. (2003) in lower redshift
($z \approx 0.1$) clusters (2/7). Since we cannot rule out that the two
hostless sources we found are not cluster SNe~Ia (i.e., that they
reside in faint background galaxies, or are not SNe at all), the lower
limit on the intergalactic SN~Ia fraction could be as low as
zero. Assuming that $1\pm1$ of these events are cluster SNe~Ia, and
that our cluster SN~Ia sample includes both the 6 likely and 7
possible events we listed above, we get a likely fraction of $1/13$
with an uncertainty of a factor of a few. This is several times below
the value we measured at lower redshift (Gal-Yam et al. 2003), as would
be expected from models (Dubinski 1998) which predict that the
intergalactic population of stars is mostly assembled between $z=1$
and $z=0$.

To summarize, we have conducted a survey for SNe in the fields of 15 
galaxy clusters, using
new and archival {\it HST}/ACS data. Each cluster was visited two to three
times, for at least one orbit. In these data, we have discovered 37
candidate transient events, of which five are likely cluster SNe~Ia,
eight are possible cluster SNe~Ia, and the rest are background or
foreground events.  We have determined cluster membership of the
candidate host galaxies using follow-up spectroscopy from ground-based
telescopes, and have measured stellar luminosities using Subaru and
{\it HST} photometry.  

We find that the SN rate in clusters at $0.5<z<0.9$, which we have
measured here accurately for the first time, is consistent with the
rates measured at lower redshifts. Our main finding is thus that there
is little or no evolution in cluster SN rates from the present time
out to $z \approx 0.9$.  The low and unevolving SN rate suggests that
an increase in ICM iron abundance between redshift $0$ and $1$, as
reported based on X-ray observations, if real, is the result of the
redistribution of iron in clusters, and not due to the production of
new iron by SNe during this period.  Two of the candidate events are
possible hostless cluster SNe~Ia, which we have discussed in the
context of the few known examples of such intergalactic SNe, and their
fraction in clusters.

In a forthcoming paper, we will combine our current result with
previous measurements to analyze the cluster SN~Ia rate as a function
of redshift, to examine the clues it can provide regarding the
progenitors of SNe~Ia, and to investigate in more detail the role of
SNe~Ia in the metal enrichment of the ICM.

Finally, although the result presented in this paper is the most
accurate cluster rate to date at high redshifts, it still suffers from
uncertainties due to the small number of SNe on which it is based, and
the difficulty in acquiring spectroscopic confirmation for SNe at such
redshifts. While spectroscopy will remain challenging in the
foreseeable future, upcoming surveys yielding larger numbers of
cluster SNe could lessen the current Poisson uncertainties in the
rates. This would lead to further progress in the study of several
issues that can be illuminated by means of cluster SN rates.

\begin{acknowledgments}

We thank the anonymous referee for very useful comments that greatly 
improved the paper. 
We wish to thank D. Poznanski and M. Sullivan for useful discussions
and help with the Keck observations.  K.S. acknowledges support from
the Kersten Visiting Fellowship Fund, and thanks the Department of
Astronomy and Astrophysics at the University of Chicago for their
hospitality during the time some of this research was conducted.  A.G.
and J.P.K. acknowledge support by the grant 07AST-F9 from the Ministry
of Science, Culture \& Sport, Israel, \& the Ministry of Research,
France.  A.G. is also supported by the Israeli Science Foundation, an
EU Seventh Framework Programme Marie Curie IRG fellowship and the
Benoziyo Center for Astrophysics, a research grant from the Peter and
Patricia Gruber Awards, and the William Z. and Eda Bess Novick New
Scientists Fund at the Weizmann Institute.  D.M. acknowledges support
by the Israel Science Foundation.  This research was supported by
National Science Foundation grants AST--0607485 and AST--0908886 to
A.V.F., as well as by NASA/{\it HST} grant GO-10793 from the Space
Telescope Science Institute (STScI), which is operated by AURA, Inc.,
under NASA contract NAS 5-26555. A.V.F. is also grateful for Department
of Energy grant DE-FG02-08ER41563.  The authors wish to recognize and
acknowledge the very significant cultural role and reverence that the
summit of Mauna Kea (where the Subaru and Keck data were obtained) has
always had within the indigenous Hawaiian community; we are most
fortunate to have the opportunity to conduct observations from this
mountain.

\end{acknowledgments}

%


\end{document}